\documentclass[pdflatex,sn-mathphys-num]{sn-jnl}

\usepackage{graphicx}%
\usepackage{multirow}%
\usepackage{amsmath,amssymb,amsfonts}%
\usepackage{amsthm}%
\usepackage{mathrsfs}%
\usepackage[title]{appendix}%
\usepackage{xcolor}%
\usepackage{textcomp}%
\usepackage{manyfoot}%
\usepackage{booktabs}%
\usepackage{algorithm}%
\usepackage{algorithmicx}%
\usepackage{algpseudocode}%
\usepackage{listings}%

\theoremstyle{thmstyleone}%
\theoremstyle{thmstyletwo}%

\theoremstyle{thmstylethree}%

\usepackage{tikz-feynman}
\usepackage{mathtools} 
\usepackage{amsmath}

\begin{document}

\title{Pion off-shell form factors
}


\author*[1,2]{\fnm{S. G.} \sur{Bondarenko}}

\author*[2,1]{\fnm{M. K.} \sur{Slautin}}
\email{slautin@jinr.ru}

\affil[1]{ \orgname{ Bogoliubov Laboratory of Theoretical Physics, JINR}, \orgaddress{ \city{Dubna}, \postcode{141980 },  \country{Russia}}}

\affil[2]{ \orgname{Dubna State University}, \orgaddress{ \city{Dubna}, \postcode{141980 }, \country{Russia}}}

\abstract{In the paper, the electromagnetic off-shell pion form factors in the Bethe-Salpeter formalism with a separable kernel are considered. Different types of vertex functions of a pion are investigated. The separable kernel of the quark-antiquark interaction is used to obtain an analytical solution of the equation. The pion constants and the form factors on both the on-shell and off-shell surfaces are calculated. The differential cross section of the reaction  $^1H(e,e' \pi^ +) n$ is also calculated in the paper. All the obtained results  are compared with experimental data. The fulfillment of the Ward-Takahashi identity for the off-shell form factors $F_1(Q^2,t)$ and $F_2(Q^2,t)$ of a pion is verified.}




\maketitle






\section{Introduction}
\label{intro}
The pion, the simplest bound system of a quark and an antiquark, occupies a special place among the mesons. Its relatively low mass—much smaller than that of other mesons—makes it a key element in the description of nuclear dynamical processes. There are many models for describing the pion: QCD sum rules~\cite{Nesterenko:1982gc};  nonrelativistic potential model~\cite{Godfrey:1985xj};
a relativistic model using the light-front formalism~\cite{Jacob:1988as}; the Nambu-Jona-Lasinio models
~\cite{Nambu:1961tp,Zhang:2024dhs,Zhang:2025iiw,Anikin:1995cf,Bernard:1986ti,Hatsuda:1985ey};
a model based on chiral symmetry~\cite{Gross:1991te}; an instanton pion
model~\cite{Anikin:2000rq}; lattice calculations ~\cite{ExtendedTwistedMass:2023hin},~\cite{Gerardin:2019vio};
recent models based on the Bethe-Salpeter equation with
dressed quark and gluon propagators~\cite{Maris:2000sk,Kekez:2020vfh,Hernandez-Pinto:2023yin}; the nonlocal models~\cite{Friesen:2025fhg}.

In~\cite{Ito:1991pv,Itzykson:1980rh}, the model based on the Bethe-Salpeter equation with the separable kernel of the quark-antiquark interaction is proposed. To calculate the elastic on-shell pion form factor, an interaction current is introduced, which in combination with the relativistic impulse approximation ensures gauge invariance of the amplitude. 

The electromagnetic (EM) form factor of the pion denoted as $F_\pi(Q^2)$ is defined on the mass shell and characterizes the spatial distribution of charge inside the pion as a function of the squared four-momentum transfer ($q^2=-Q^2$). At low $q^2$, this form factor can be directly accessed through elastic scattering experiments of pions on electrons. However, at intermediate and high values of the squared momentum transfer  $q^2$, direct extraction becomes difficult because the short lifetime of the pion makes it unsuitable for such measurements.

An experimental alternative for accessing the form factor in these regions is provided by the exclusive Sullivan process~\cite{Sullivan:1971kd}. This method involves analyzing the cross sections of the electromagnetic reaction $^1H(e,e' \pi^ +) n$. However, since the pion in the intermediate state is virtual, the extracted form factor is half-off-shell, which requires further theoretical interpretation.

In~\cite{Choi:2019nvk,Leao:2024agy}, the space-like off-shell pion form factors are investigated in the BS approach within the relativistic impulse approximation. In addition, the off-shell pion form factor was extracted from Jefferson Lab
data for pion electroproduction on proton~\cite{JeffersonLab:2008gyl,JeffersonLabFpi-2:2006ysh}.

In the paper, using a model based on the relativistic covariant Bethe-Salpeter equation with a separable kernel, the pion static properties, transition and elastic EM pion form factors (both on- and half-off-shell)  are investigated. 
The choice of this model is due to the simplicity of the analytical solution  of the vertex function of the pion.
The Ward-Takahashi identity~\cite{Ward:1950xp,Takahashi:1957xn} is numerically verified for the off-shell pion form factors $F_1$ and $F_2$,
and  the importance of the interaction current is shown. The longitudinal cross section $\sigma_L$ is calculated using the form factor $F_1$~\cite{JeffersonLab:2008gyl}~\cite{JeffersonLab:2008jve}. The paper also studies the dependence of the obtained off-shell form factors and static properties of the pion on the types of kernel functions. The obtained results are also compared with experimental data.

The paper is organized as follows: after introduction (Sec.1), the Bethe-Salpeter approach for the pion is given (Sec.2). The details of the pion static properties and on-shell form factor calculations are described in Sec.3. The off-shell elastic form factors are discussed in Sec.5. The results of the calculations and discussion are given in Sec.5, a short summary is in Sec.6.

\section{Bethe-Salpeter approach}
\label{sec:1}

The Bethe-Salpeter equation for the vertex function of the pion is written as follows~\cite{Ito:1991sz}:
\begin{equation}
\Gamma_{\alpha\beta}(k;p)=i\int\frac{d^4k''}{(2\pi)^4}V_{\alpha\beta:\epsilon\lambda}(k,k'';p)S_{\lambda\eta}(k''+p/2)\Gamma_{\eta\zeta}(k'';p)S_{\zeta\epsilon}(k''-p/2).
\label{eqgamma}
\end{equation}
where $p,k$ are the total and relative 4-momenta of the pion, respectively ($p=k_1+k_2$ and $k=(k_1-k_2)/2$), and $V(k',k;p)$ is the interaction kernel. 
The pion mass $m_{\pi}$ is defined on the pion mass shell as $p^2=m_\pi^2$, where $m_\pi=140$ MeV.
The quark propagator with mass $m$ has the form
$S(k)=(\hat{k}-m+i\epsilon)^{-1}$. The Greek symbols denote the Dirac indices.

In this paper, a separable interaction kernel of rank one is considered
in the following form:
 \begin{equation}
     V_{\alpha\beta:\delta\gamma}(k',k;p)=\gamma_{\alpha\beta}^5 f(k'^2) \times \gamma_{\delta\gamma}^5 f(k^2).
 \end{equation}

For simplicity, in the paper, only the $k^2$ dependence of the scalar function is considered. In this case, the solution for the vertex function~(\ref{eqgamma}) of the charged pion can be written in the following form:
\begin{equation}
    \Gamma(k;p)\equiv\Gamma(k)={N\gamma^5}{f(k^2)},
\end{equation}
where $N$ is the constant that corresponds to the normalization of the EM current.

The radial part of the vertex function is chosen in the following form:
\begin{eqnarray}
&&f(k^2)=({k^2-\Lambda^2+i\epsilon})^{-1} \textrm{--- monopole},\nonumber\\
&&f(k^2)=({(k^2-\Lambda^2)^2+i\epsilon})^{-1} \textrm{--- dipole},
\end{eqnarray}
where the parameter $\Lambda$ is related to the size of the pion.

\section{Pion constants and transition $\gamma^*\pi^0\rightarrow{}\gamma$ form factor}

To fix the model parameters $m, \Lambda$, the observables $f_{\pi} , r_{\pi\gamma}, <r^2_{\pi}>$
are considered.

By writing down the amplitude of the weak decay of the pion according to the Feynman diagram, the following formula can be obtained
for the decay constant of the pion $f_{\pi}$:
\begin{equation}
    f_{\pi}={-i4m}\sqrt{N_c}N\int\frac{d^4k}{(2\pi)^4}\frac{f(k^2)}{([k-p/2]^2-m^2+i\epsilon)([k+p/2]^2-m^2+i\epsilon)},
\end{equation}
where $N_c=3$  is the number of quark colors.

The two-photon decay of the pion is expressed by the following integral:
\begin{eqnarray}
&&G_{\pi\gamma}(p^2,q^2_1,q^2_2)=\frac{4\sqrt{2}Nm}{\sqrt{N_c}}\times\\
&&\int\frac{d^4k}{(2\pi^4)}\frac{f(k^2)}{((k+p/2)^2-m^2+i\epsilon)((k-p/2)^2-m^2+i\epsilon)((k+(q_1-q_2)/2)^2-m^2+i\epsilon)}.
\nonumber
\end{eqnarray}

The decay width is determined by the transition form factor at zero:
\begin{equation}
    \Gamma_{\pi^0\rightarrow{}\gamma\gamma}=\frac{\pi\alpha^2}{4}m_{\pi}^3G^2_{\pi\gamma}(m^2_{\pi},0,0),
\end{equation}
where $\alpha={e^2}/{4\pi}={1}/{137}$  is the fine structure constant.

The process of $\gamma^*\pi^0\rightarrow{}\gamma$ transition can be considered as a cross channel for the two-photon decay of the pion and can be written as:
\begin{equation}
F_{\pi\gamma}(Q^2)=G_{\pi\gamma}(m^2_{\pi},Q^2,0).
\label{eqffpigamma}
\end{equation}

The radius of the junction $\gamma^*\pi^0\rightarrow{}\gamma$ is defined as
\begin{equation}
    <r^2_{\pi\gamma}>=-6\frac{F'_{\pi\gamma}(Q^2)}{F_{\pi\gamma}(Q^2)}\bigg|_{Q^2=0},
\end{equation}
where $F'_{\pi\gamma}$ is the derivative of the transition form factor (\ref{eqffpigamma}) at zero.

The charge radius of the pion is defined as the derivative of the form factor at zero transmitted momentum:
\begin{equation}
<r_\pi^2>=-6\frac{d}{dQ^2} F_{\pi}(Q^2)\bigg|_{Q^2=0},
\end{equation}
where $F_{\pi}(Q^2)$ is discussed in the next section.

\section{Elastic EM pion form factors}

In the section, the elastic EM pion form factors are considered.
The photon-pion vertex $G_\mu$ can generally be represented in the following form~\cite{Rudy:1994qb}:
\begin{equation}
    G_\mu(p,p')=(p'+p)_{\mu}F_1(q^2,p^2,p'^{2})+q_{\mu}F_2(q^2,p^2,p'^{2}),
\label{gammapi}
\end{equation}
where $p, p'$ are the initial and final 4-momenta, $q=p'-p$ is the transferred 4-momentum
of the virtual photon at the vertex. The Ward-Takahashi identity is satisfied by this vertex on the off-shell~\cite{Rudy:1994qb,Itzykson:1980rh}:
\begin{equation}\label{WTI}
    q^{\mu}G_{\mu}(p,p')=\Delta^{-1}(p')-\Delta^{-1}(p),
\end{equation}
where
\begin{equation}
    \Delta(p)=\frac{1}{p^2-m_{\pi}^2-\Pi(p^2)+i\varepsilon}
\end{equation}
is a fully renormalized propagator, and the renormalized
self-energy of the pion $\Pi(p^ 2)$ is bounded by the condition on the mass shell: $\Pi(m_{\pi}^ 2)=0$.

In the case when the initial pion is off-shell 
and the final pion is on the mass shell,
it follows from equations~(\ref{gammapi}),(\ref{WTI}) that
\begin{equation}\label{WTIF2}
    F_2(Q^2,t)=\frac{t-m_\pi^2}{Q^2}[F_1(0,t)-F_1(Q^2,t)],
\end{equation}
where
$t=p^2$ is the off-shell parameter.
The dependence on $p'^2=m_\pi^2$ is omitted for the half-off-shell form factors below.

It should be noted that when the pion is completely on the mass 
shell ($p^2=p'^2=m_\pi^2$), then $F_2(Q^2,m_\pi^2)=0$,
and it does not contribute to normalization of the form factors $F_1(0,m_\pi^2)=F_\pi(0)=1$. 
It also ensures the conservation of the EM current.

The function $g$ can be determined using~(\ref{WTIF2})
\begin{equation}\label{WTIg}
    g(Q^2,t)\equiv \frac{F_2(Q^2,t)}{t-m_\pi^2},
\end{equation}
which does not turn to zero on the mass shell. 
The value of $g(0,m^2_p)$, as shown in equations (\ref{WTIF2}) and (\ref{WTIg}), is related to the charge radius of the pion:
\begin{equation}
    g(0,m_\pi^2)=-\frac{\partial}{\partial Q^2}F_1(Q^2,m_\pi^2)\bigg|_{Q^2=0}=\frac{1}{6}\langle r_\pi^2\rangle.
\end{equation}

The new form factor $g(Q^2,t)$ is a physical quantity that can be observed in experiments.

The EM pion current consists of two parts - relativistic impulse approximation (RIA) and interaction (INT) terms:
\begin{equation}
    F_{(1,2)}(q^2,t)=F^{\textrm{RIA}}_{(1,2)}(q^2,t)+F^{\textrm{INT}}_{(1,2)}(q^2,t).
\end{equation}

The first part corresponds to the interaction of the virtual photon with the pion constituents - the quarks:
\begin{eqnarray}
&&\langle{J}^{\mu}_{\textrm{RIA}}\rangle=
\nonumber\\
&&ie_1\int\frac{d^4k}{(2\pi)^4}\{\Gamma(k^2)S(p/2+k+q)\gamma^\mu S(p/2+k)\Gamma((k+q/2)^2)S(p/2-k)\}
\nonumber\\
&&+(1\longleftrightarrow2),
\nonumber\\
&&
\label{eqfriatot}
\end{eqnarray}
while the second part describes the interaction of the photon with the nonlocal quark-antiquark field
\begin{eqnarray}
&&\langle{J}^{\mu}_{\textrm{INT}}\rangle=
\nonumber \\
&&-ie_1\int\frac{d^4k}{(2\pi)^4}\frac{(k+q/4)^{\mu}}{(k+q/4)\cdot{q}}\{\left[\overline{\Gamma}(k+q/2)-\overline{\Gamma}(k)\right]S(k+p/2)\Gamma(k){S}(k-p/2)\}
\nonumber \\
 &&+ie_1\int\frac{d^4k}{(2\pi)^4}\frac{(k-q/4)^{\mu}}{(k-q/4)\cdot{q}}\{S(k-(p+q)/2)\overline{\Gamma}(k)S(k+(p+q)/2)\left[\Gamma(k-q/2)-\Gamma(k)\right]\}
\nonumber\\
&&+(1\longleftrightarrow2).
\label{eqfinttot}
\end{eqnarray}
Equation (\ref{eqfinttot}) is obtained in~\cite{Gross:1991te} by the minimal substitution method in the kernel $V(k,k')$.

In the case of  on-shell,  in (\ref{eqfriatot}) and (\ref{eqfinttot}) the pion momentum squares are equal to
$p^2=t, p'^2=m_\pi^2$, where the variable $t$ is the off-shell parameter. The on-shell case is defined as $t=m_\pi^2$.

The integrals (\ref{eqfriatot}) and (\ref{eqfinttot}) were calculated by several methods: 
using the Wick-rotation procedure calculating the residues with the Cauchy theorem, and 
using the Feynman parameter method; details can be found in the papers~\cite{Bondarenko:2025aep,Bondarenko:2025qch}.  All methods give the same results.

To calculate the interaction part, the following analytic expressions were used:
\begin{eqnarray}
&& \left[{\Gamma}(k+q/2)-{\Gamma}(k)\right] \to\\
&&\mbox{monopole: }\nonumber\\
&&\gamma_5 [\frac{1}{((k+q/2)^2-\Lambda^2)}-\frac{1}{(k^2-\Lambda^2)}] =
\gamma_5\frac{-(k+q/4)\cdot q}{((k+q/2)^2-\Lambda^2)(k^2-\Lambda^2)}\nonumber\\
&&\mbox{dipole: }\nonumber\\
&&\gamma_5 [\frac{1}{((k+q/2)^2-\Lambda^2)^2}-\frac{1}{(k^2-\Lambda^2)^2}] =
\gamma_5\frac{(k+q/4)\cdot q (-2k^2-k\cdot q-q^2/4+2\Lambda^2)}{((k+q/2)^2-\Lambda^2)^2(k^2-\Lambda^2)^2}\nonumber
\end{eqnarray}
The factor $(k+q/4)\cdot q$ in the numerator cancels out the same factor in the denominator. The second term in (\ref{eqfinttot}) can be treated in a similar way. The obtained integrals are typical integrals to be calculated
using the Feynman parameter method. The master integrals for the numerators with $k_\mu$ up to the 5th power are given in Appendix~\ref{feynformulas}.

\section{Results and discussion}
\label{sec:4}

\begin{table}[h]

\caption{Table of the model parameters $m, \Lambda$ and observables $f_{\pi} , r_{\pi\gamma}, <r^2_{\pi}>$, $\Gamma_{\pi^0 \rightarrow{} \gamma\gamma}$, $g(0,m_\pi^2)$}
\label{tab:1}      

\begin{tabular}{llllllll}
\hline\noalign{\smallskip}
& \scriptsize {$m$ } & \scriptsize{$\Lambda$ } & \scriptsize{$f_{\pi}$ } & \scriptsize{$r_{\pi\gamma}$ } & \scriptsize{$<r^2_{\pi}>$} & \scriptsize{$\Gamma_{\pi^0\rightarrow{}\gamma\gamma}$ } & \scriptsize{$g(0,m_\pi^2)$ } \\

& \scriptsize (MeV) & \scriptsize (MeV)& \scriptsize (MeV)& \scriptsize (fm)& \scriptsize ($\textrm{fm}^2$)  & \scriptsize (eV) & \scriptsize $(\textrm{GeV})^{-2}$\\
\noalign{\smallskip}\hline\noalign{\smallskip}
\scriptsize monopole & \scriptsize260 & \scriptsize550 & \scriptsize143.363(1)  & \scriptsize0.640(1)  & \scriptsize0.459(1)  & \scriptsize7.25(1) & \scriptsize 1.97(2) \\  
\scriptsize  & \scriptsize & \scriptsize & \scriptsize  & \scriptsize  & \scriptsize0.466(1) -0.007(1)  & \scriptsize & \scriptsize  \\  
\scriptsize dipole & \scriptsize245 &\scriptsize 1100 & \scriptsize135.962(1)  & \scriptsize0.652(1)  & \scriptsize0.438(1)& \scriptsize7.74(1)  & 1.79(3)\\ 
\scriptsize  & \scriptsize &\scriptsize & \scriptsize  & \scriptsize  & \scriptsize0.442(1) -0.004(1) & \scriptsize  & \\ 

    
 \scriptsize Exp.~\cite{PDG} &  &  & \scriptsize 130.410(2) & \scriptsize 0.659(4) & \scriptsize 0.434(8) & \scriptsize 7.82(1) 
 & \scriptsize 1.86(5) \\  
\noalign{\smallskip}\hline
\end{tabular}
\end{table}

First, the parameters of the model $m_q, \Lambda$ are fixed by calculating the best values of the pion constants.
Table~\ref{tab:1} gives the parameters of the model and the calculated pion constants. 
For $<r^2_{\pi}>$, separate contributions of the RIA and INT part are given.
It is seen that the results obtained with the dipole model are closer to the experimental values (less than 5\%) than
with the monopole one. 


\begin{figure}[htb]
\begin{minipage}[htb]{0.5\linewidth}
{\includegraphics[width=\linewidth]{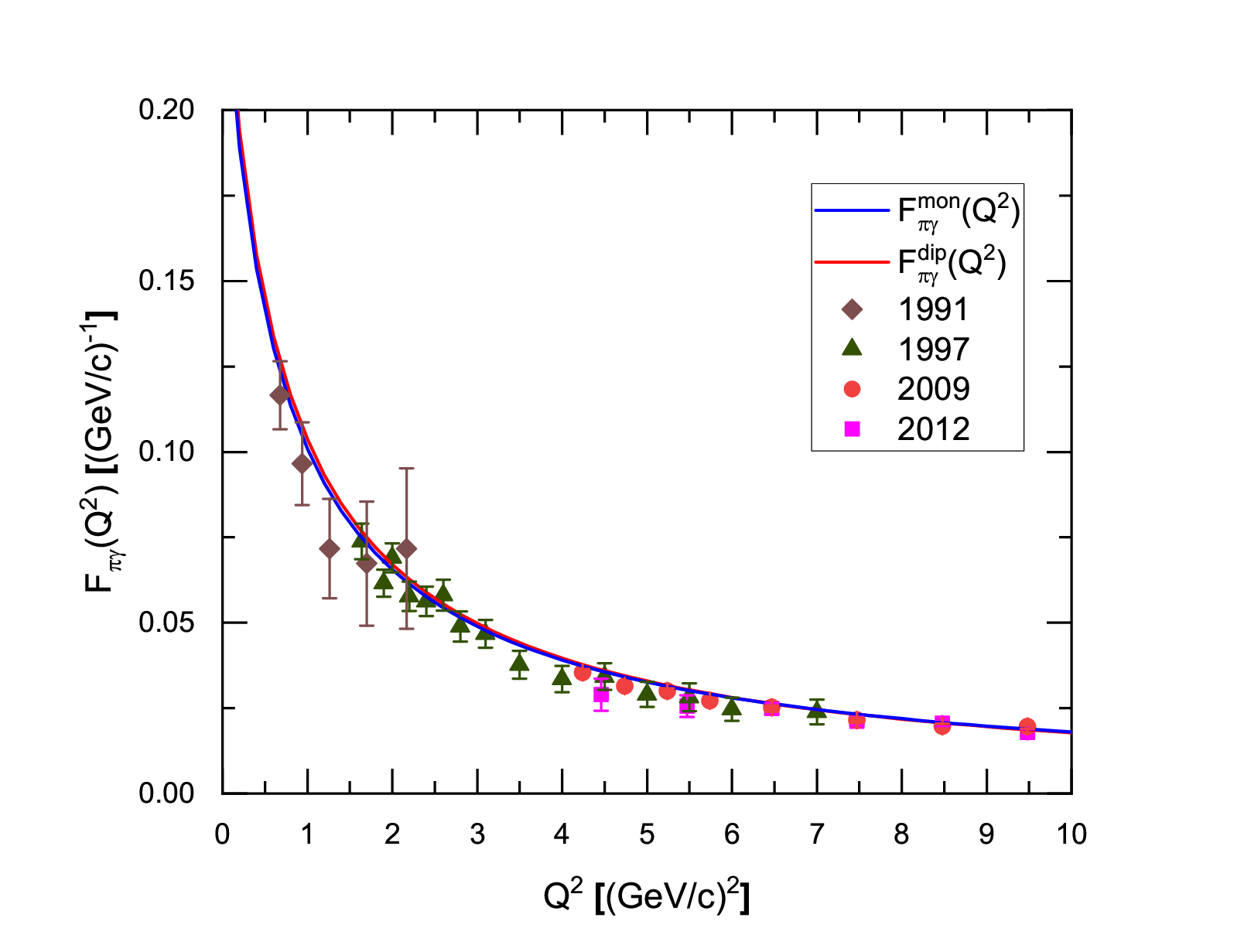}}
\end{minipage}
\begin{minipage}[htb]{0.5\linewidth}
{\includegraphics[width=\linewidth]{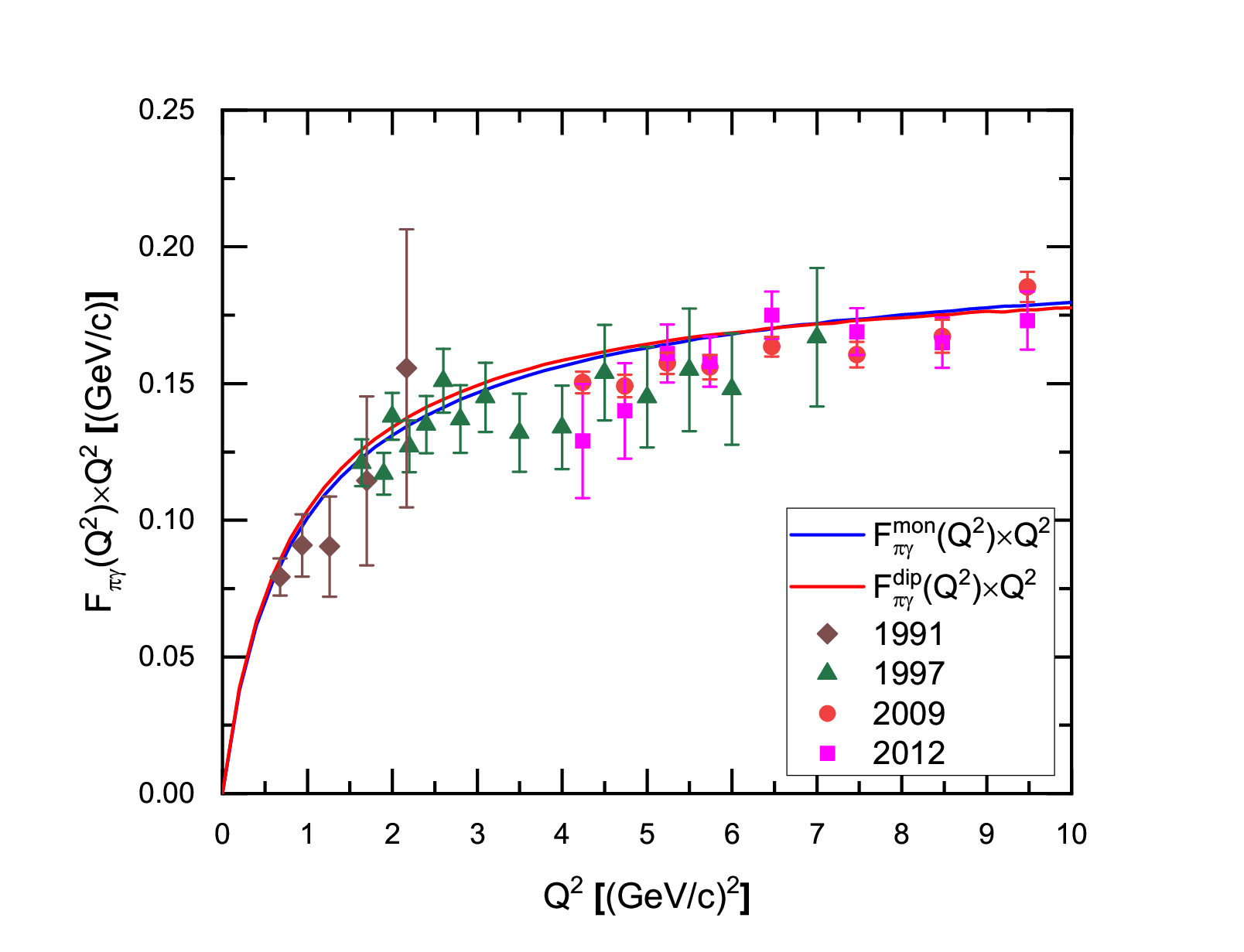}}
\end{minipage}
\caption{\label{figfpigammaexp}\small{Comparison of the calculated $F_{\pi\gamma}(Q^2)$ (left panel) and
$F_{\pi\gamma}(Q^2)\times{Q^2}$ (right panel) with experimental data. 
Rhombus --- \cite{CELLO:1990klc}, triangle --- \cite{CLEO:1997fho}, circle --- \cite{BaBar:2009rrj}, square --- \cite{Belle:2012wwz}}}
\end{figure}

The obtained parameters are used to calculate the transition form factor $\gamma^*\pi^0\rightarrow{}\gamma$. 
Figures~\ref{figfpigammaexp} 
show the results of the calculation of $F_{\pi\gamma}$ (left panel) and $F_{\pi\gamma}\times{Q^2}$ (right panel) using the monopole and dipole models in comparison with experimental data. 
Comparison of the form factor with experimental data shows good agreement for both monopole and dipole functions
with a slight difference at the $Q^2 >~$5$~GeV^2/c^2$. The dipole comes out more hollow on the asymptotics of $F_{\pi\gamma}\times{Q^2}$ than the monopole. 
The value of $F_{\pi\gamma}\times{Q^2}$ at high $Q^2$ is determined by perturbative QCD predictions, $F_{\pi\gamma}\times{Q^2} \to 2f_\pi$,
and form factor demonstrates good agreement with this limit.

\begin{figure}[htb]
\begin{minipage}[htb]{0.5\linewidth}
{\includegraphics[width=\linewidth]{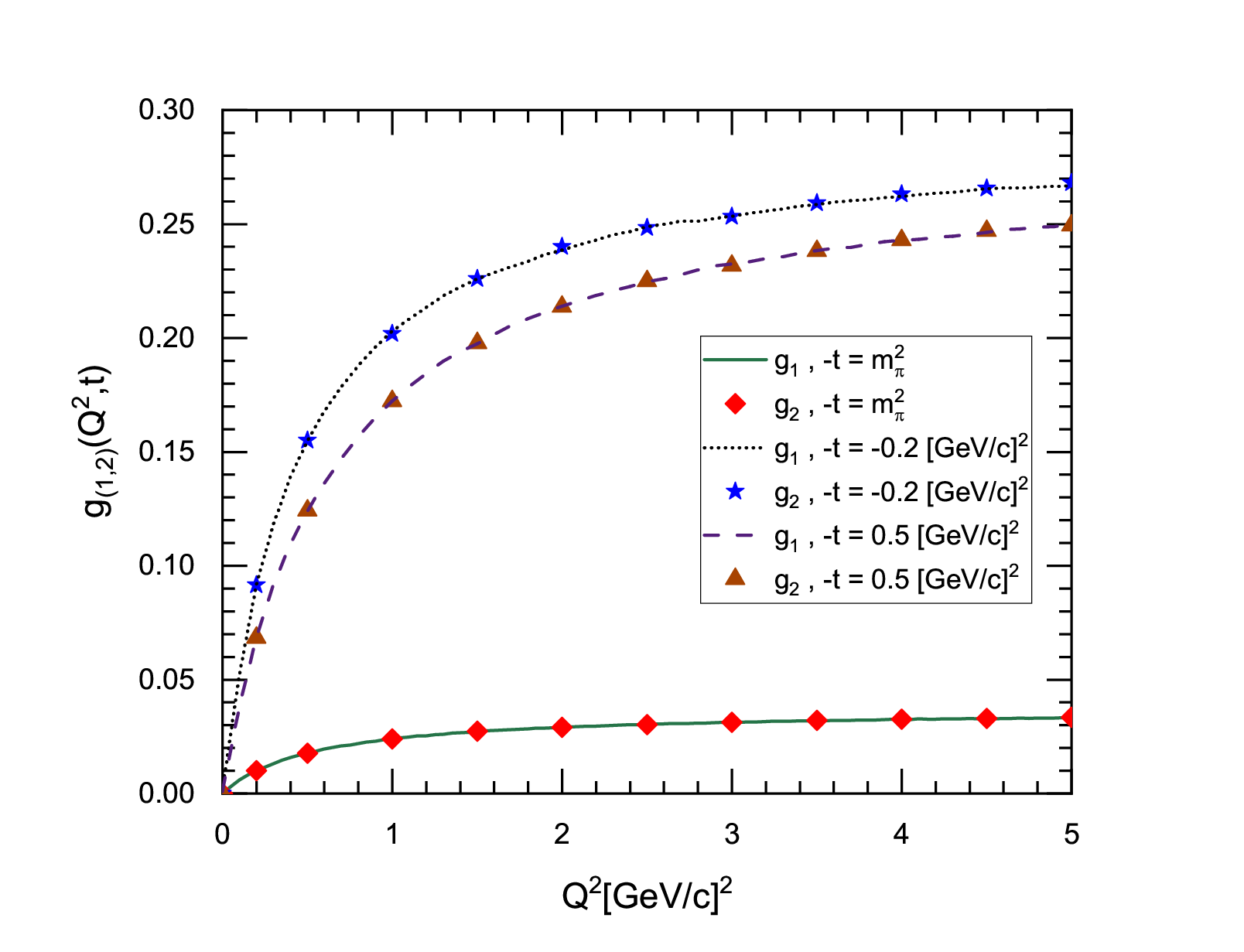}}
\caption{\label{fig_WTI} 
Numerical verification of the Ward-Takahashi identity at different $t$.}
\end{minipage}
\begin{minipage}[htb]{0.5\linewidth}
{\includegraphics[width=\linewidth]{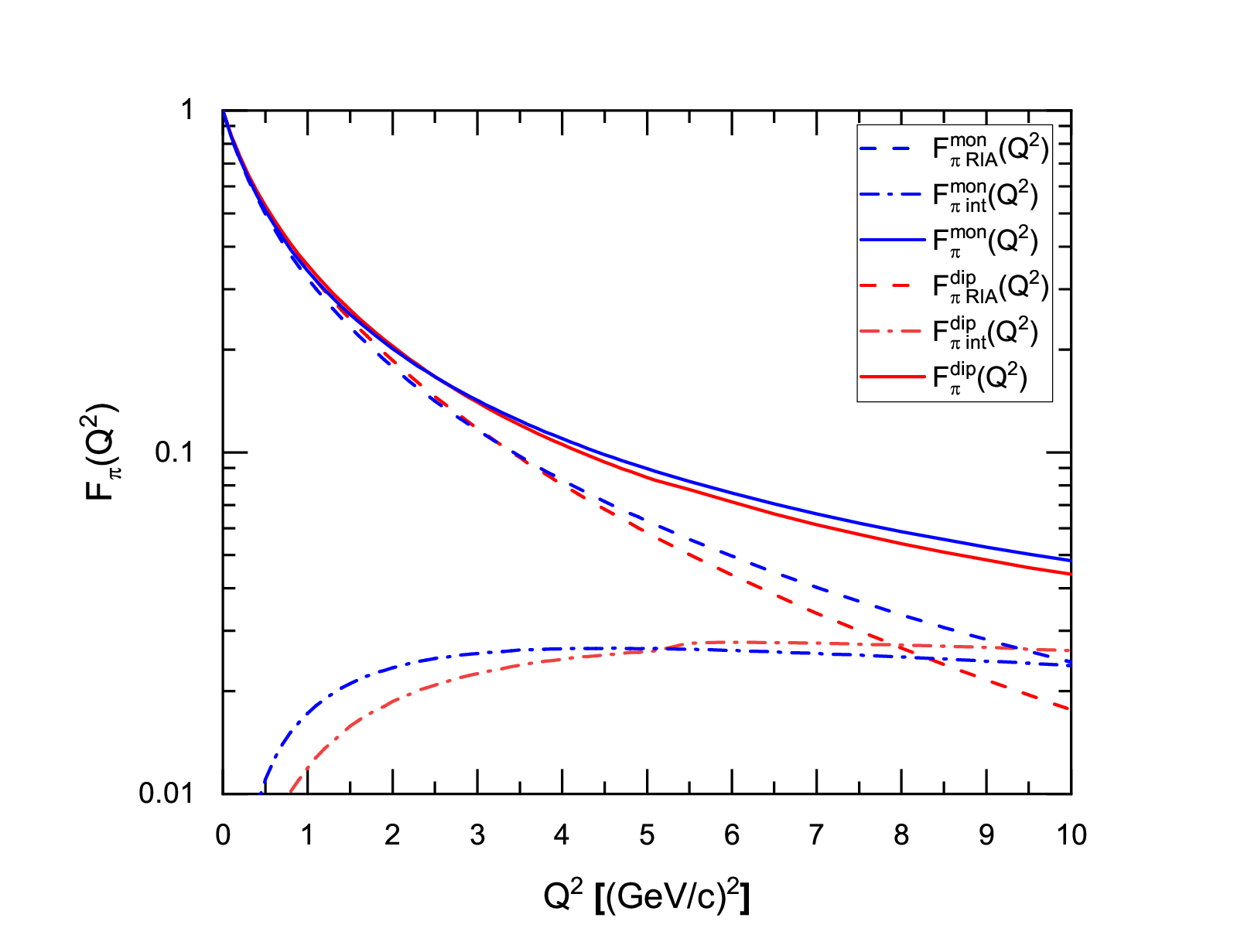}}
\caption{\label{figfriaintsum}\small{Comparison of the contribution of the relativistic impulse approximation and interaction current to the $F_\pi(Q^2)$.}}
\end{minipage}
\end{figure}

Since the Ward-Takahashi identities play an important role in the self-consistency of the model, the comparison of two types of calculation of the form factor $g(Q^2,t)$ is shown in figure~\ref{fig_WTI}.
The directly calculated (from~(\ref{WTIg})) off-shell form factor $g_2=|Q^2F_2(Q^2,t)|$  and the form factors calculated using the Ward-Takahashi identity $g_1=|(t-m_{\pi}^2)[F_1(0,t)-F_1(Q^2,t)])|$ 
for different off-shell parameters $t$ are presented. As can be seen, the Ward Takahashi identity holds for any $t$ and $Q^2$. 
Both the  RIA and INT contributions are taken into account in $F_1(Q^2,t)$, $F_2(Q^2,t)$ and the INT part  is strongly necessary to fulfill the WTI.
It should be noted that the fulfillment of the WTI gives a powerful test of the analytic and numeric calculations of the elastic form factors.



Figure~\ref{figfriaintsum} shows a comparison of the contributions of RIA and INT to the pion form factor $F_\pi(Q^2)$ for two types of vertex functions. It can be seen that the contribution of INT is significant and necessary. It should be noted that the INT contribution is positive, which differs from the results in~\cite{Ito:1991sz,Ito:1991pv}, where it is negative and smaller in magnitude.
At $Q^2=10~\textrm{GeV}^2$, the contributions are comparable to each other, and for the dipole, the INT contribution is even greater than the contribution of the RIA.

\begin{figure}[htb]
\begin{minipage}[htb]{0.5\linewidth}
{\includegraphics[width=\linewidth]{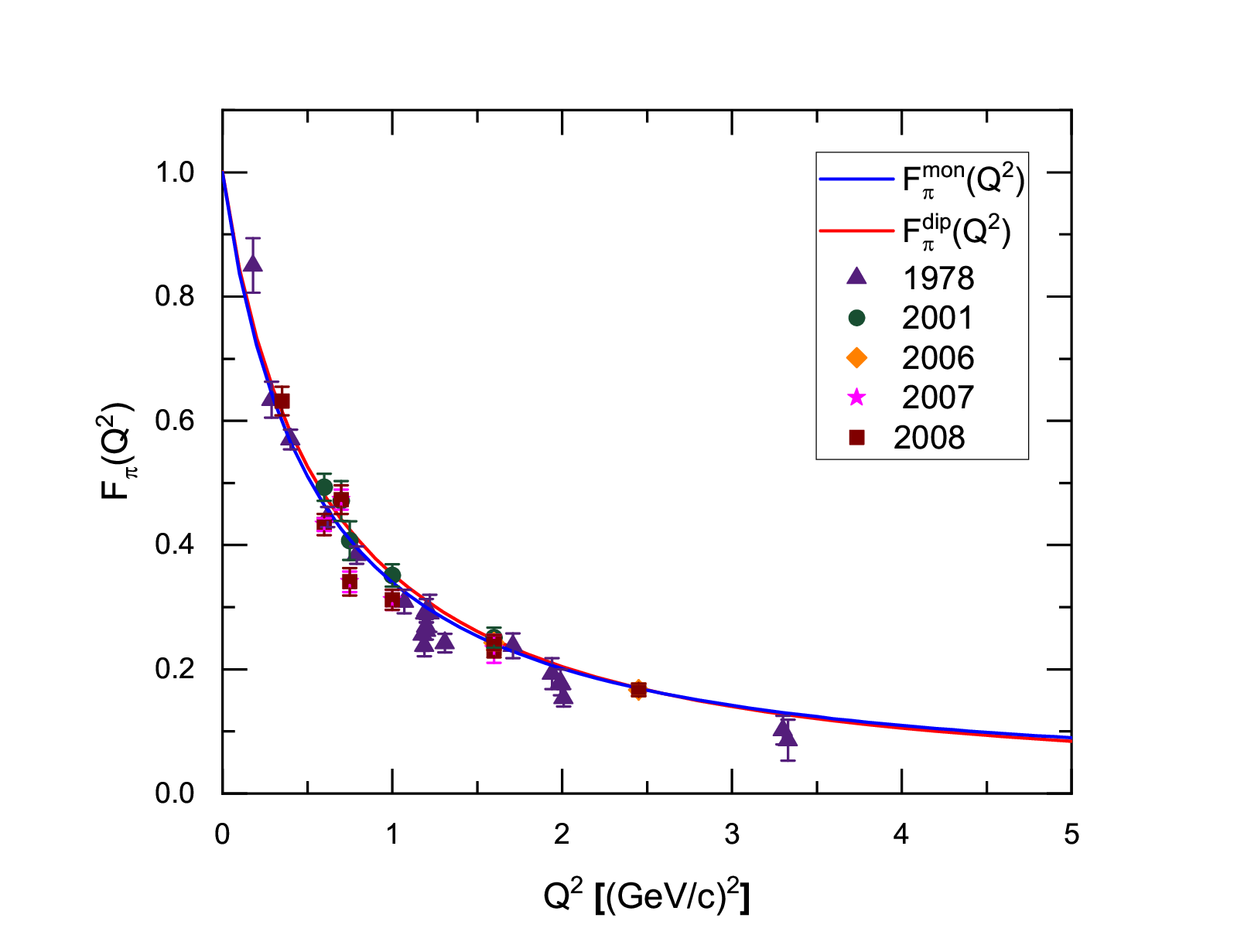}}
\end{minipage}
\begin{minipage}[htb]{0.5\linewidth}
{\includegraphics[width=\linewidth]{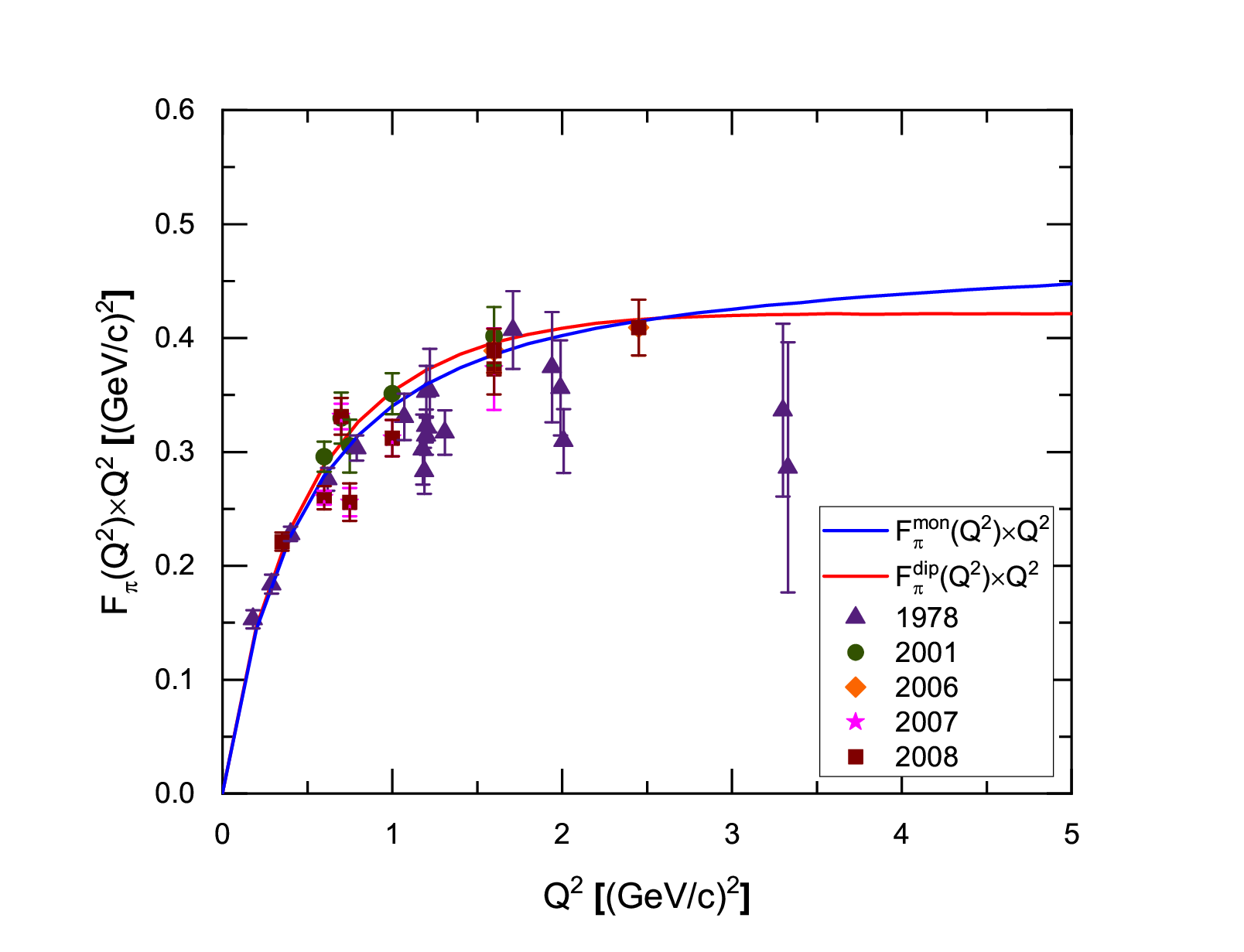}}
\end{minipage}
\caption{\label{figfexp}\small{Comparison of the calculated $F_{\pi}(Q^2)$ (left panel)
and $F_{\pi}(Q^2)\times{Q^2}$ (right panel) with experimental data.
Triangle --- \cite{Bebek:1977pe}, circle --- \cite{JeffersonLabFpi:2000nlc}, rhombus --- \cite{JeffersonLabFpi-2:2006ysh}, star --- \cite{JeffersonLabFpi:2007vir}, square --- \cite{JeffersonLab:2008jve}}}
\end{figure}

\begin{figure}[htb]
\begin{minipage}[htb]{0.5\linewidth}
{\includegraphics[width=\linewidth]{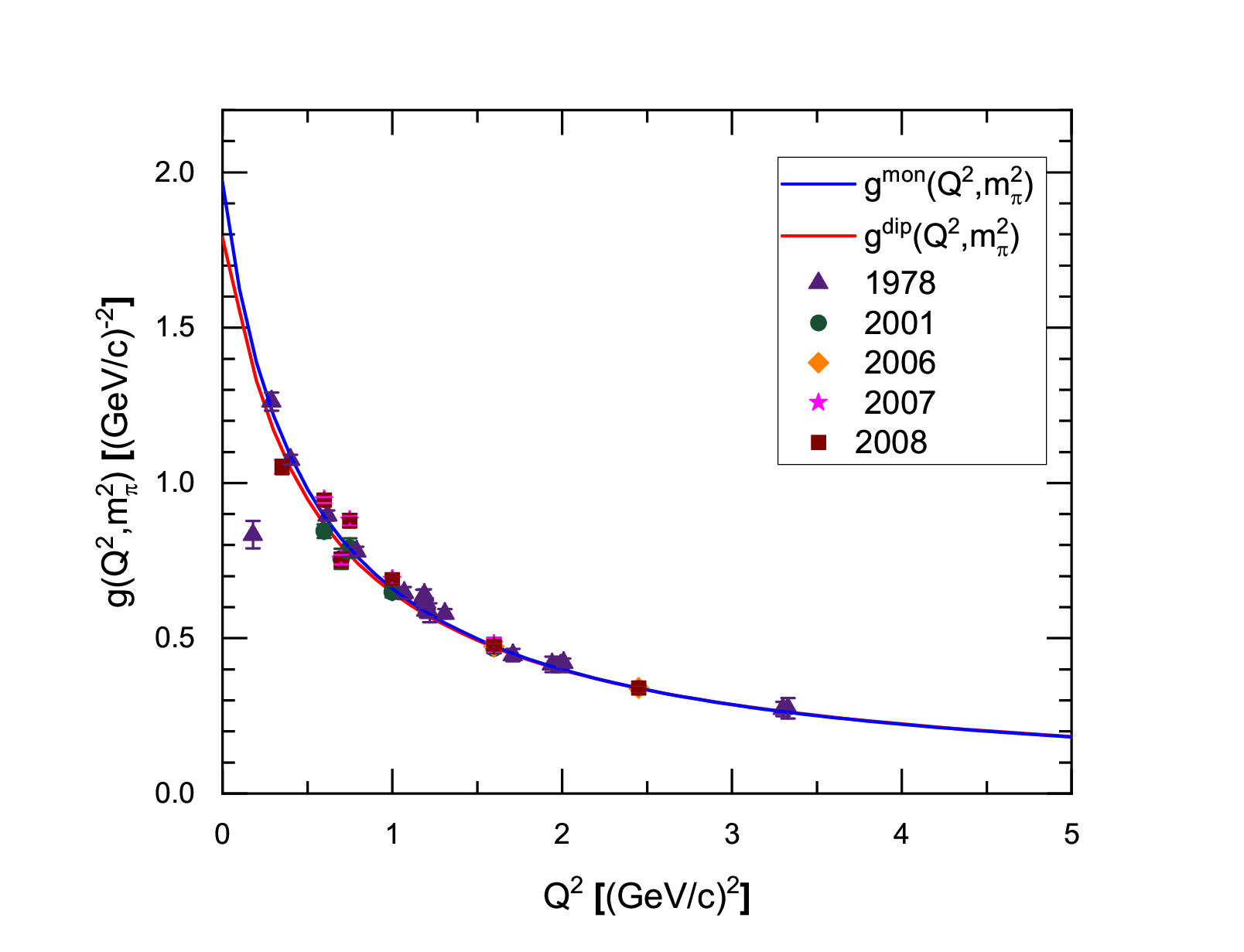}}
\end{minipage}
\begin{minipage}[htb]{0.5\linewidth}
{\includegraphics[width=\linewidth]{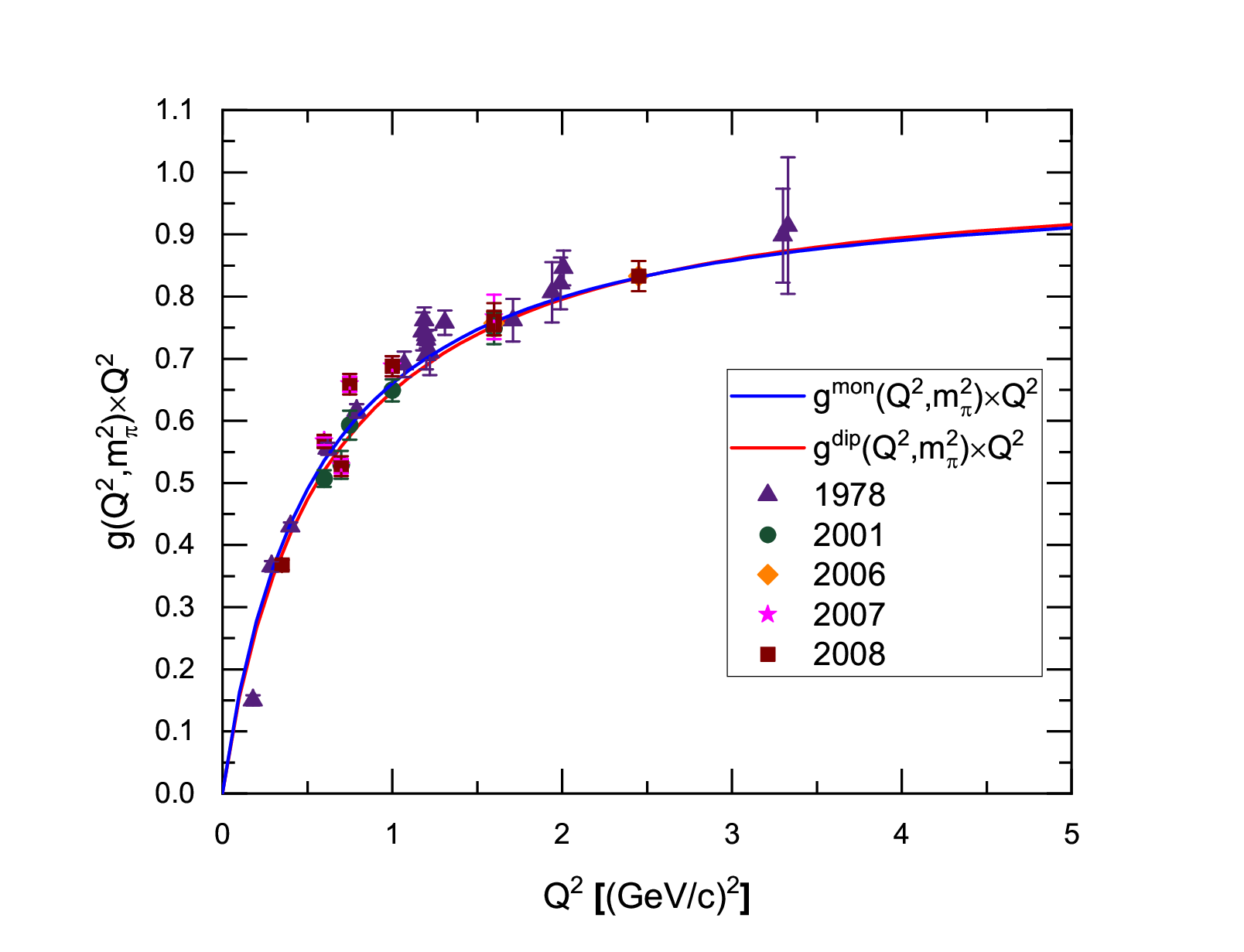}}
\end{minipage}
\caption{\label{figgexp}\small{Comparison of the calculated $g(Q^2,m_\pi^2)$ (left panel)
and $g(Q^2,m_\pi^2)\times Q^2$ (right panel) with experimental data (as in figure~\ref{figfexp}).
}}
\end{figure}

Figures ~\ref{figfexp} and ~\ref{figgexp} show the results for the elastic electromagnetic form factors $F_{\pi}$ (left panel) and $F_{\pi}\times{Q^2}$, $g(Q^2,m_\pi)$ (left panel) and  $g(Q^2,m_\pi^2)\times Q^2$ (right panel) calculated 
using the monopole and dipole models in comparison with experimental data, respectively.
Both models demonstrate a good description of experimental data.
It can also be noted that for the form factor $g(Q^2,m_\pi^2)$ practically does not depend on the choice of the pion vertex function. The experimental data 
$g(Q^2,m_\pi^2)$ are calculated with the help of~(\ref{WTIF2}) and the experimental data for $F_\pi(Q^2)$.


\begin{figure}[htb]
\begin{minipage}[htb]{0.5\linewidth}
{\includegraphics[width=\linewidth]{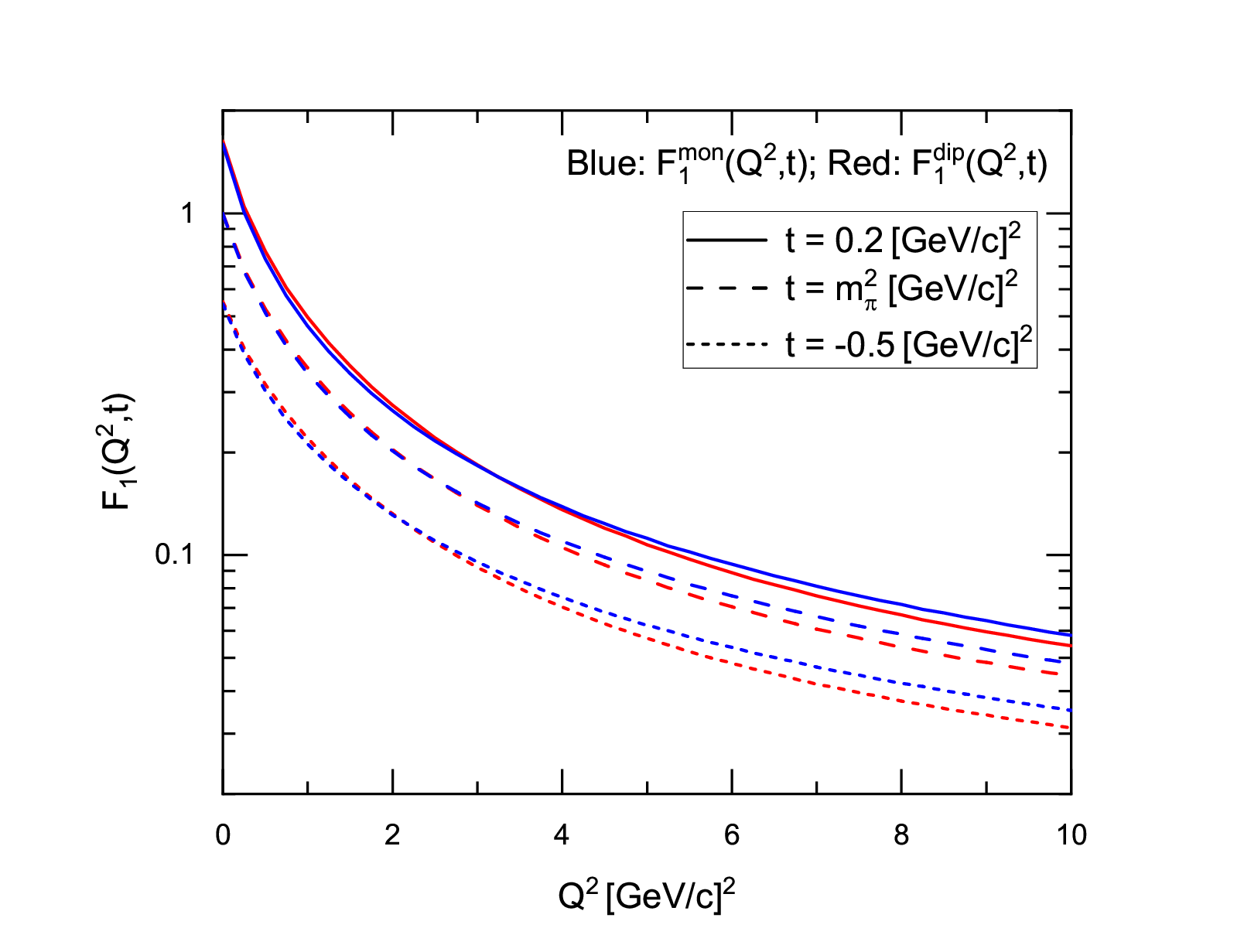}}
\end{minipage}
\begin{minipage}[htb]{0.5\linewidth}
{\includegraphics[width=\linewidth]{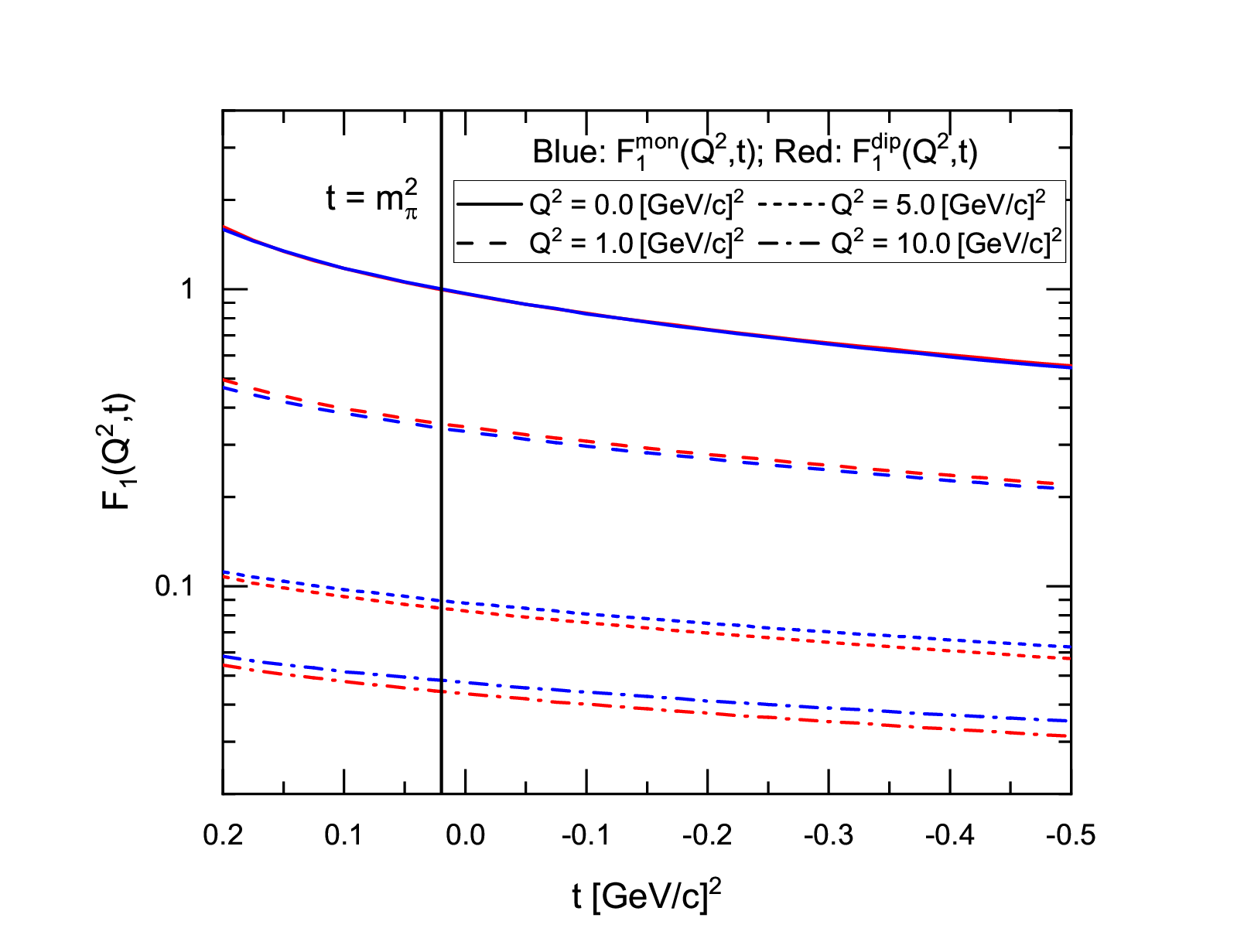}}
\end{minipage}
\caption{\label{figf1q2t}\small{Off-shell form factor $F_1(Q^2,t)$ as a function
of $Q^2$ at fixed $t$ (left panel) and of $t$ at fixed $Q^2$ (right panel).
}}
\end{figure}


\begin{figure}[htb]
\begin{minipage}[htb]{0.5\linewidth}
{\includegraphics[width=\linewidth]{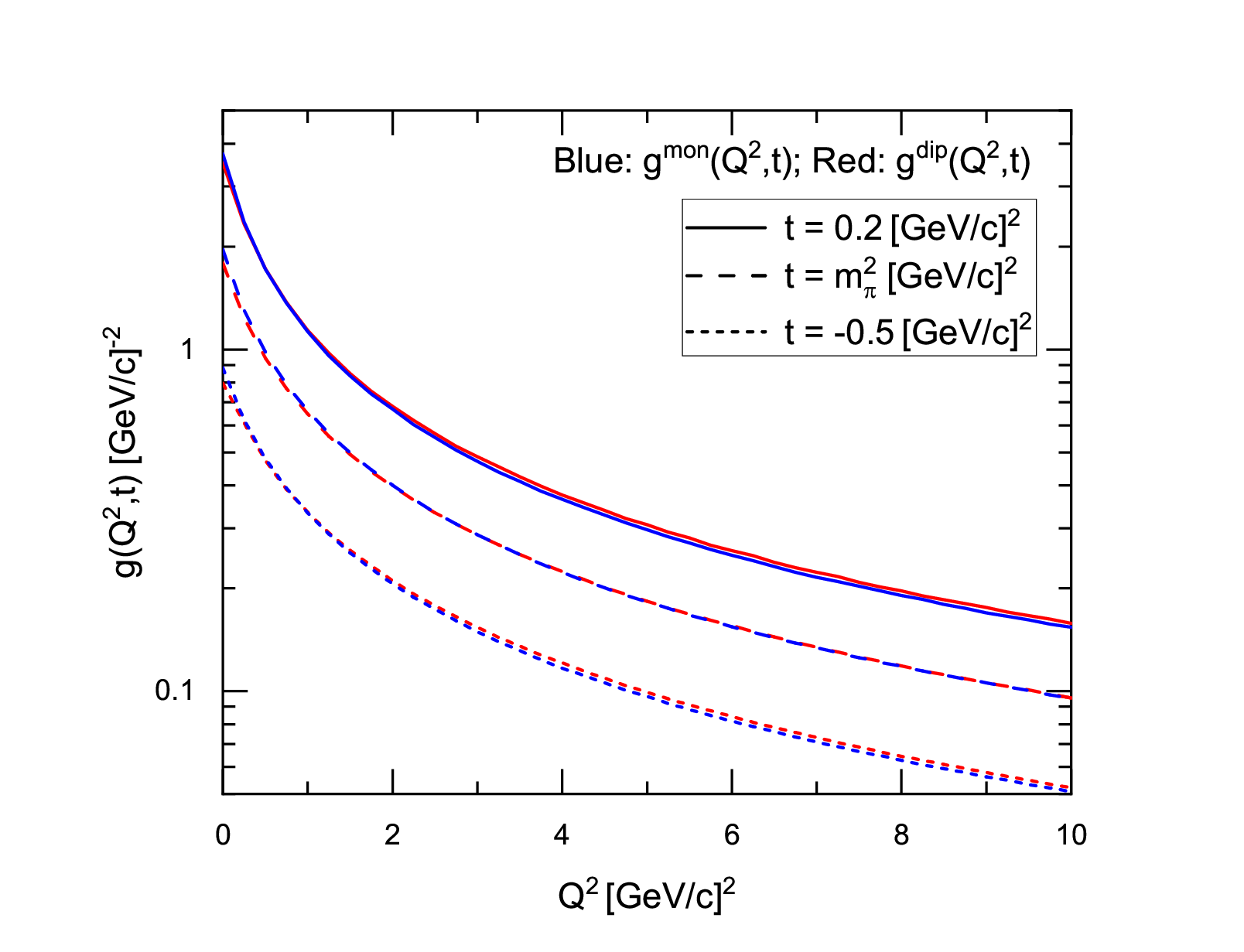}}
\end{minipage}
\begin{minipage}[htb]{0.5\linewidth}
{\includegraphics[width=\linewidth]{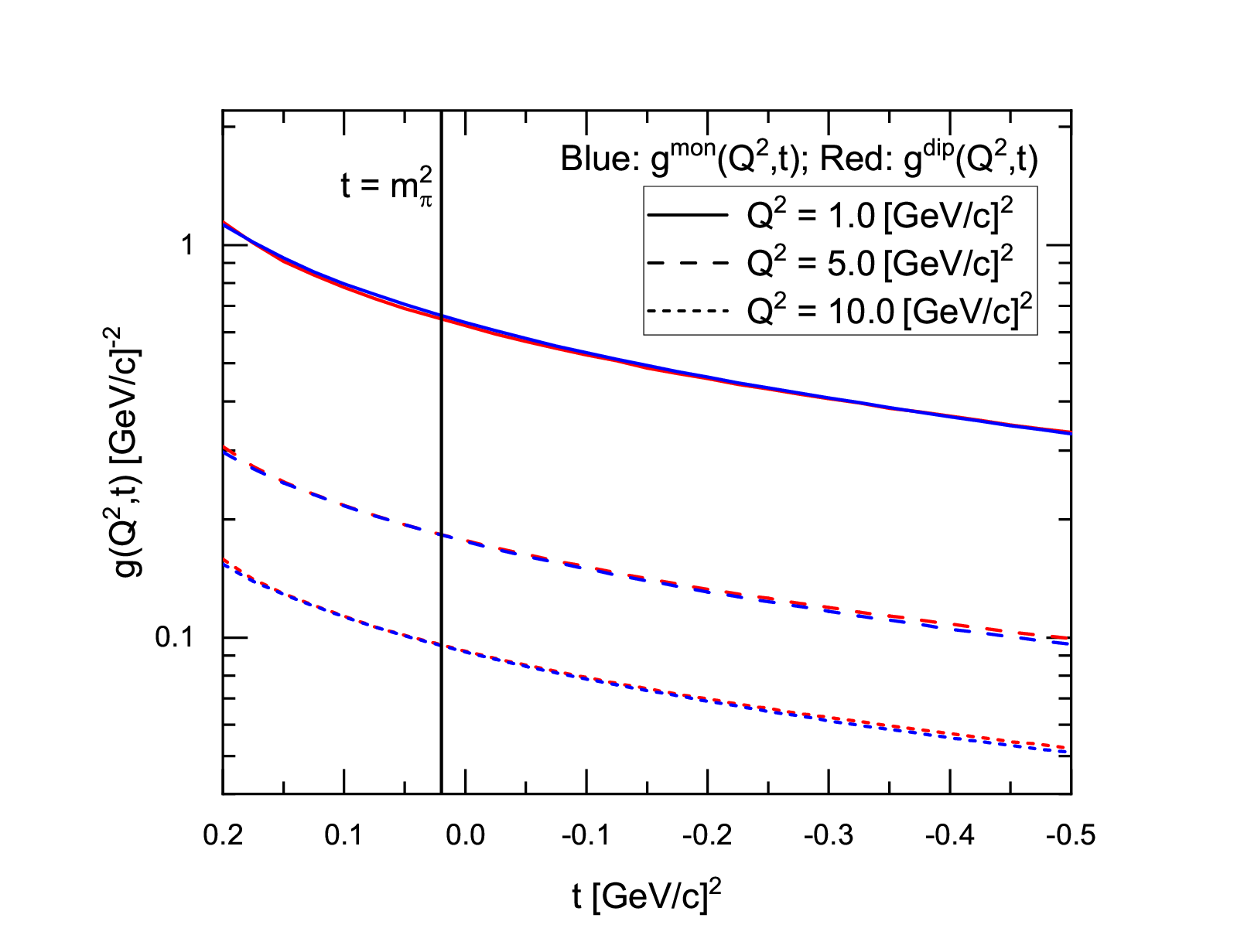}}
\end{minipage}
\caption{\label{figgq2t}\small{Off-shell form factor $g(Q^2,t)$ as a function
of $Q^2$ at fixed $t$ (left panel) and of $t$ at fixed $Q^2$ (right panel).
}}
\end{figure}

On the off-shell surface, the form factors $F_1(Q^2,t)$ and $g(Q^2,t)$ depend on two parameters: the square of the transmitted momentum $Q^2$ and the off-shell parameter $t$.
Figures~\ref{figf1q2t} show the off-shell form factor $F_1(Q^2,t)$ as a function
of $Q^2$ at fixed $t$ (left panel) and of $t$ at fixed $Q^2$ (right panel). 
It is seen that two models (monopole and dipole) differ in the range of momentum transfer squared
$Q^2 > 3$ GeV/c$^2$, where the dipole is smaller in magnitude compared to the monopole.

The same is shown in figures~\ref{figgq2t} but for the off-shell form factor $g(Q^2,t)$. There is
only a small difference for two types of models for $g(Q^2,t)$.

\begin{figure}[htb]
\begin{minipage}[htb]{0.5\linewidth}
{\includegraphics[width=\linewidth]{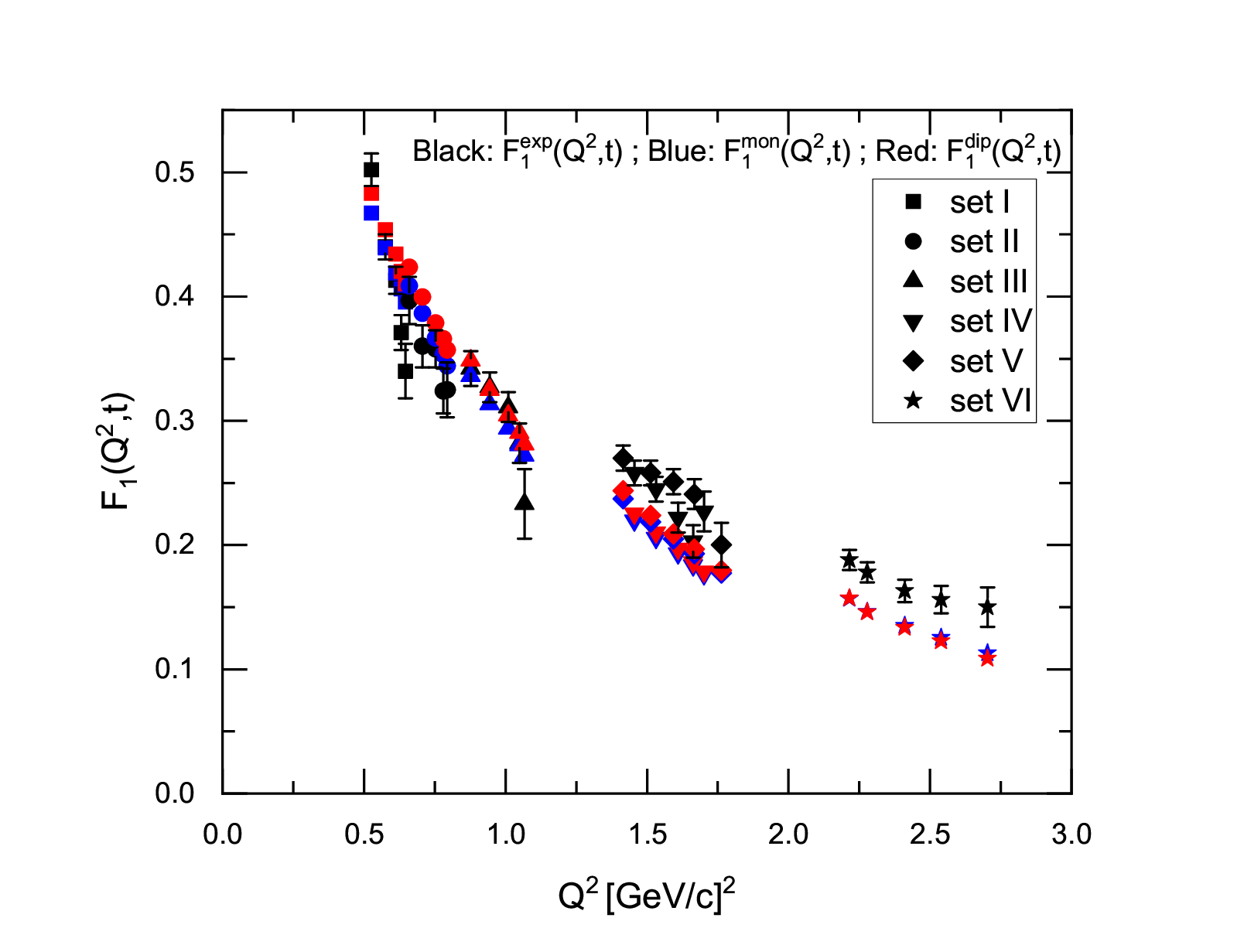}}
\end{minipage}
\begin{minipage}[htb]{0.5\linewidth}
{\includegraphics[width=\linewidth]{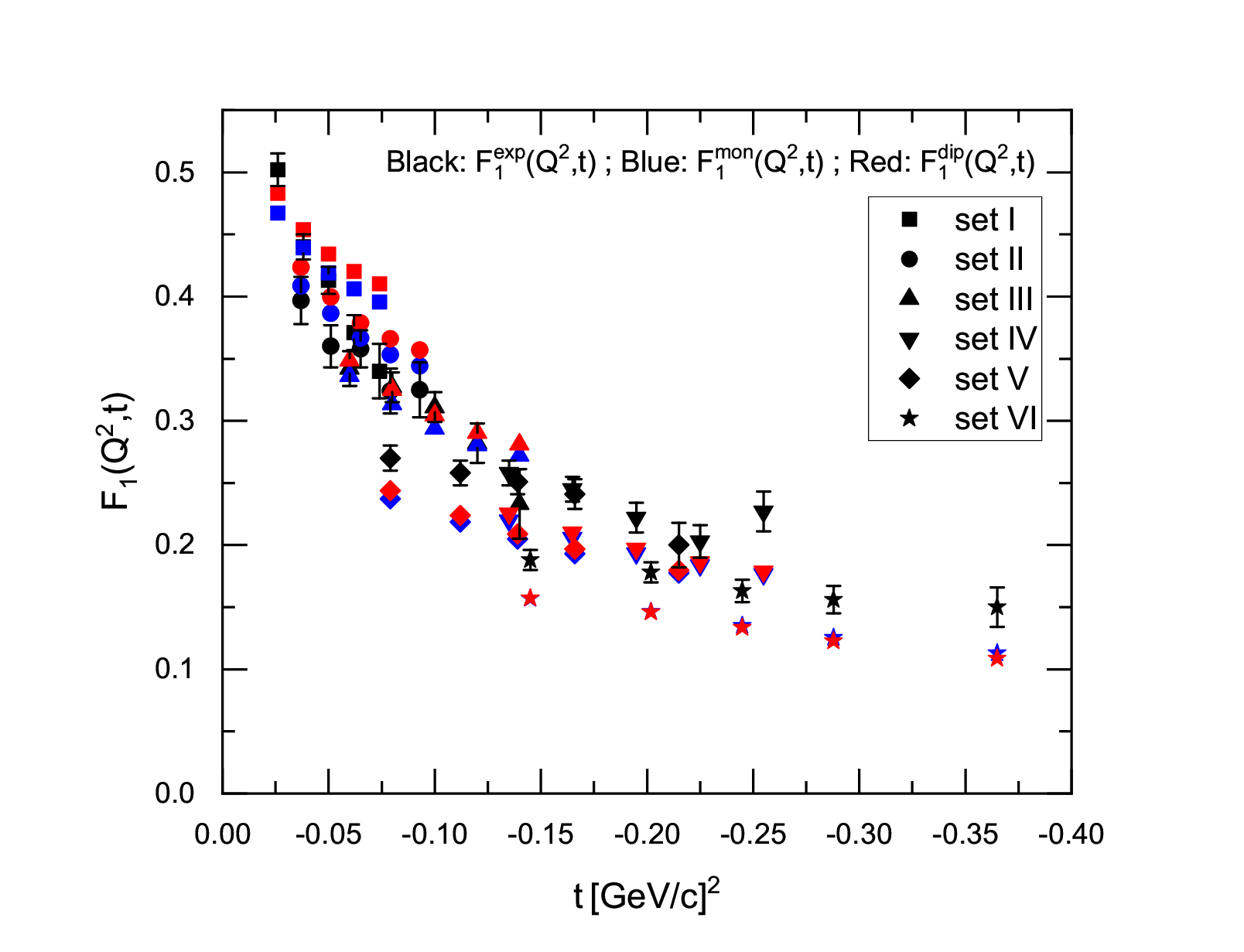}}
\end{minipage}
\caption{\label{figF1off}
The half-off-shell form factor $F_1(Q^2,t)$ as a function of $Q^2$ (left panel) and $t$ (right panel).}
\end{figure}

\begin{figure}[htb]
\begin{minipage}[htb]{0.5\linewidth}
{\includegraphics[width=\linewidth]{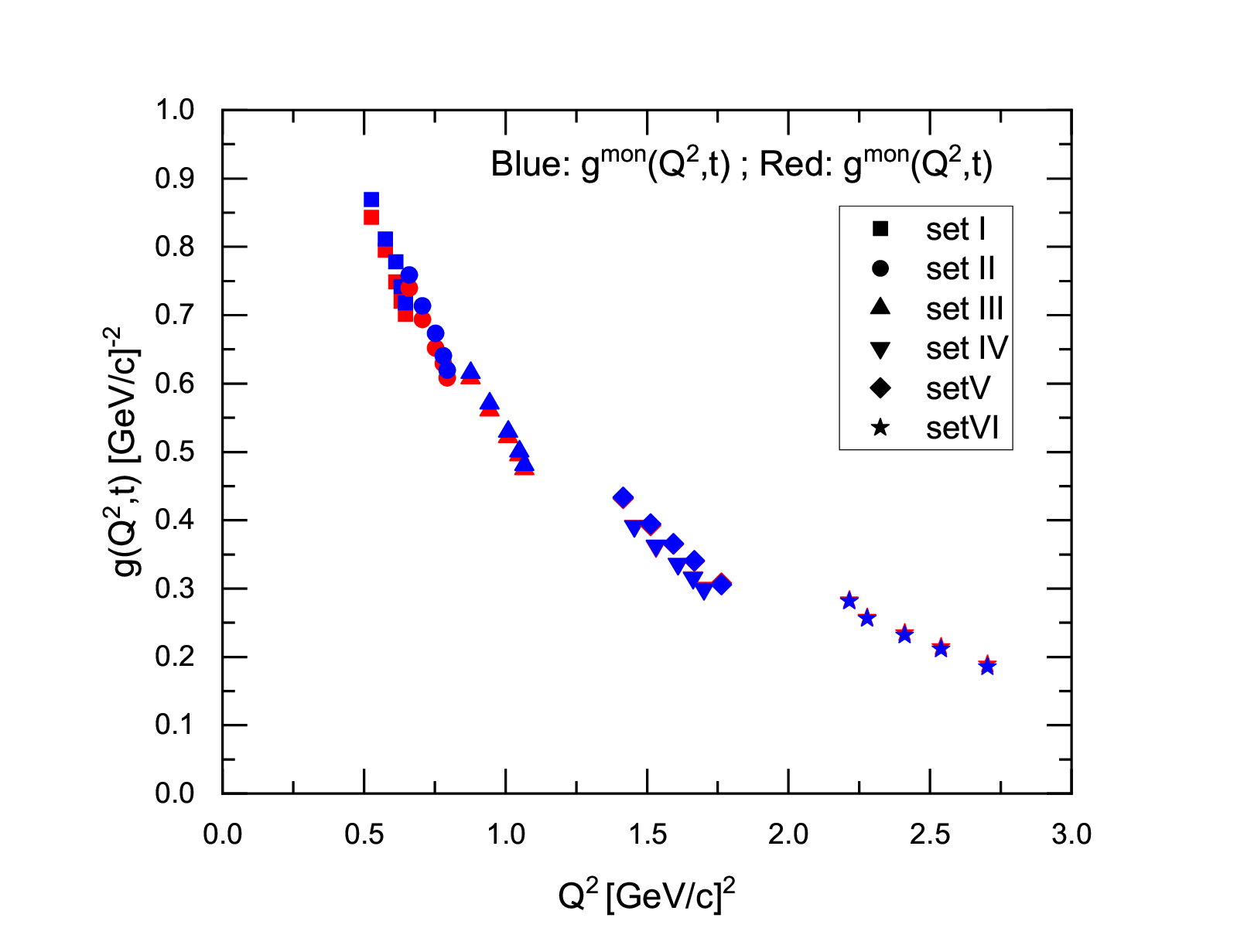}}
\end{minipage}
\begin{minipage}[htb]{0.5\linewidth}
{\includegraphics[width=\linewidth]{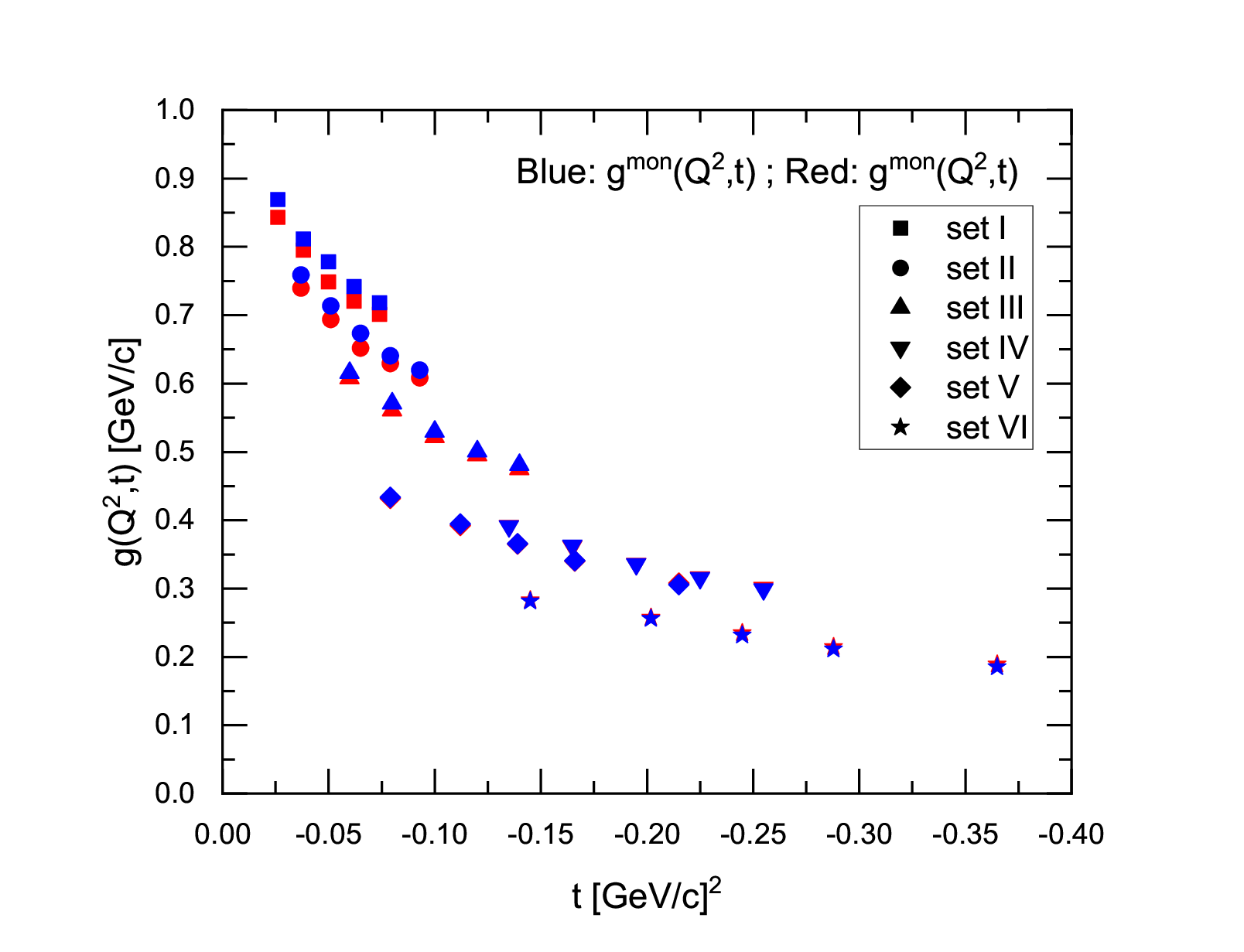}}
\end{minipage}
\caption{\label{figgoff}
The half-off-shell form factor $g(Q^2,t)$ as a function of $Q^2$ (left panel) and $t$ (right panel).}
\end{figure}

Figures \ref{figF1off} show a comparison of the calculated $F_1(Q^2,t)$ form factors in monopole and dipole forms with experimental data obtained from the longitudinal differential cross section  in~\cite{Choi:2019nvk,Leao:2024agy} assuming
that ${d\sigma_{\textrm{\tiny L}}}/{dt} \sim |F_1(Q^2,t)|^2$.
Every point is calculated at the experimental values of
$<Q^2>, t$ combined into six sets.
The results are presented in the following sets~\cite{JeffersonLab:2008gyl}: Set I: $<Q^2>=0.60~\textrm{GeV}^2, W=1.95~\textrm{GeV}$; Set II: $<Q^2>=0.75~\textrm{GeV}^2, W=1.95 ~\textrm{GeV}$; Set III: $<Q^2>=1.00~\textrm{GeV}^2, W=1.95~\textrm{GeV}$ ; Set IV: $<Q^2>=1.60~\textrm{GeV}^2, W=1.95~\textrm{GeV}$; Set V: $<Q^2>=1.60~\textrm{GeV}^2, W=2.22 ~\textrm{GeV}$;  Set VI: $<Q^2>=2.45~\textrm{GeV}^2, W=2.22~\textrm{GeV}$. As can be seen in the graph, the form factor $F_1(Q^2,t)$ is weakly dependent on the choice of the pion vertex function. The experimental data are not described exactly, but the calculated form factors have the same behavior as the experimental data for any $Q^2$ and $t$. Figures \ref{figgoff} show the calculated form factor $g(Q^2,t)$ for two types of vertex functions. The form factors are similar to each other for any $Q^2$ and $t$, as in the case of the form factor $F_1(Q^2,t)$.

\begin{figure}[htb]
\begin{minipage}[htb]{0.5\linewidth}
{\includegraphics[width=\linewidth]{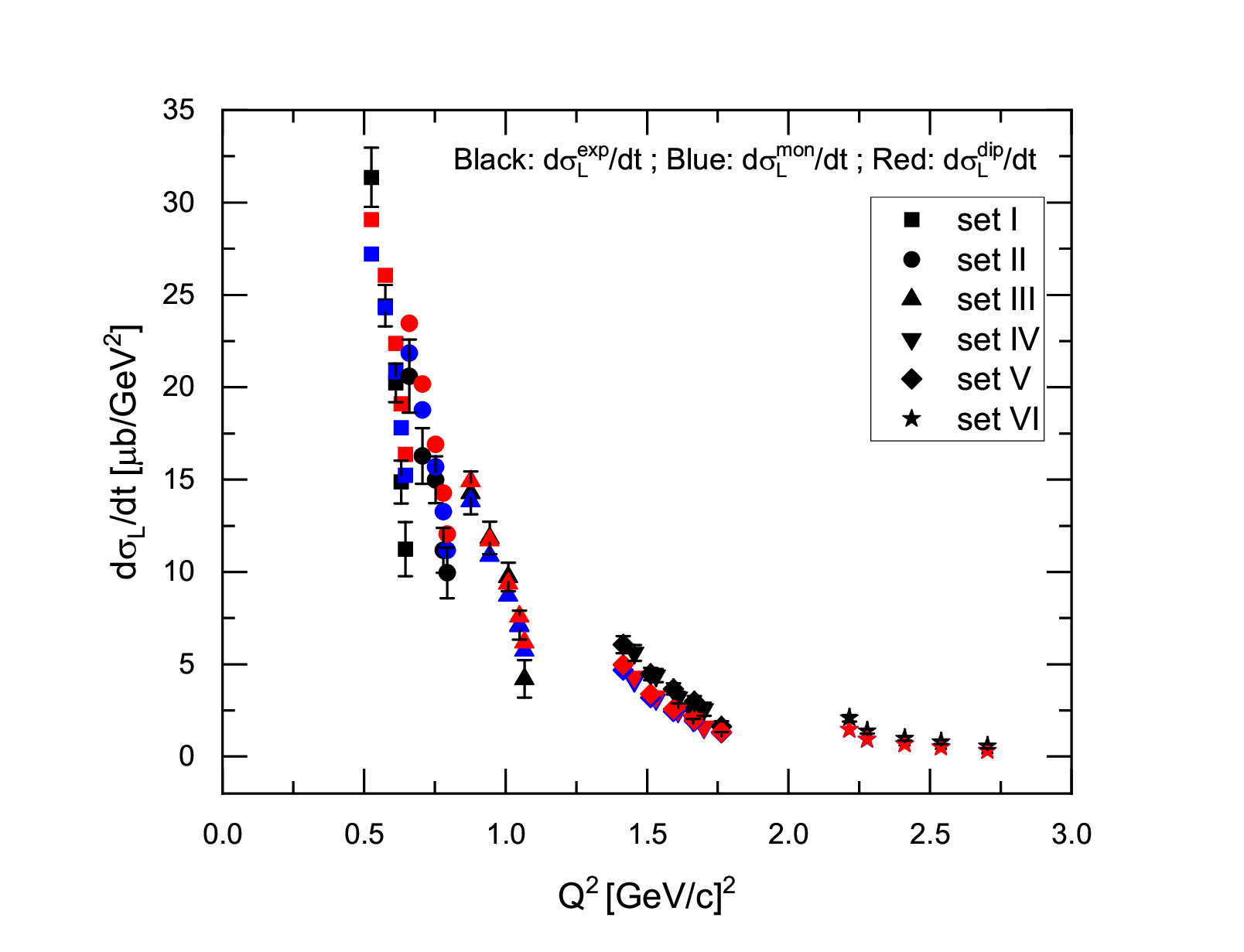}}
\end{minipage}
\begin{minipage}[htb]{0.5\linewidth}
{\includegraphics[width=\linewidth]{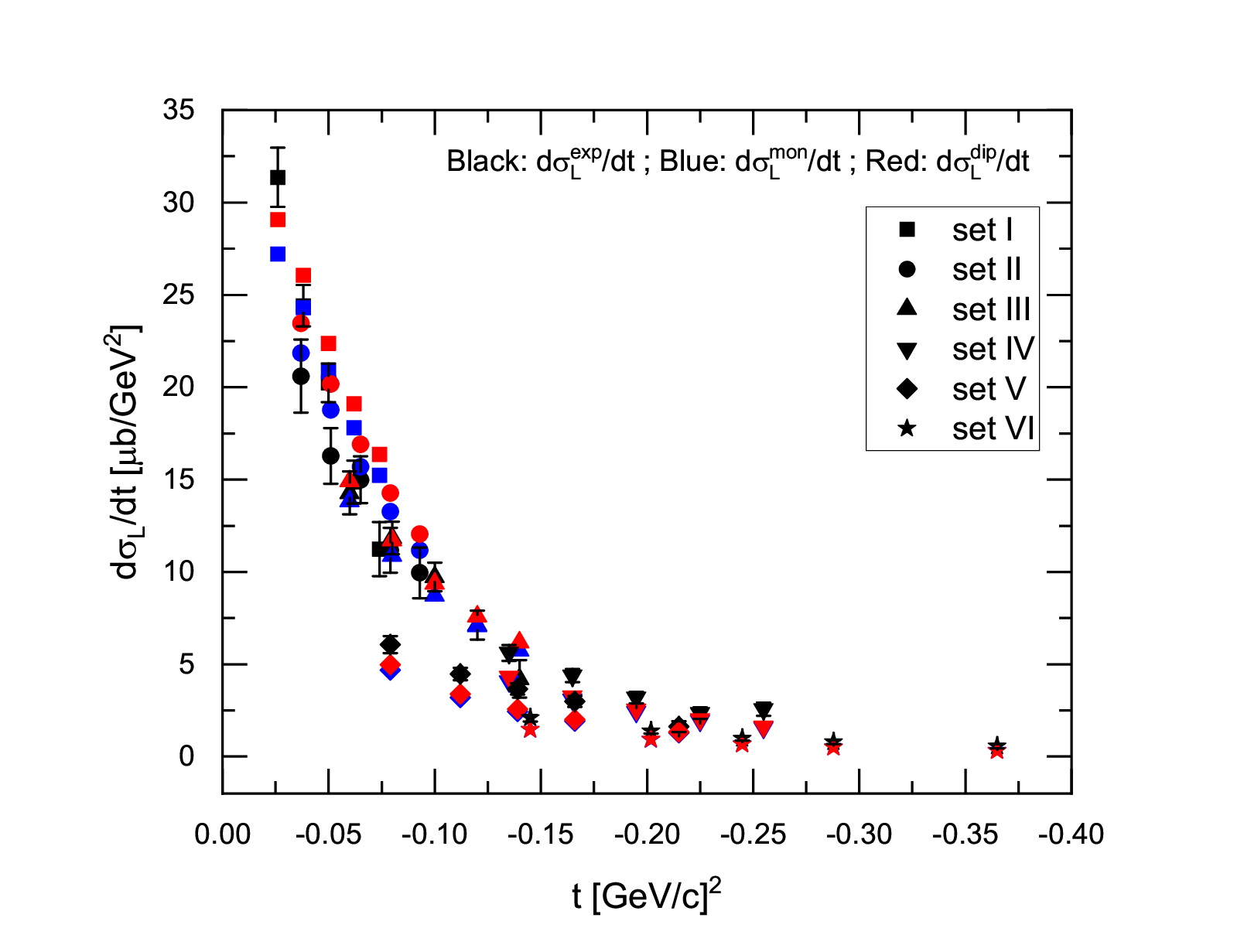}}
\end{minipage}
\caption{\label{fig_cross}
The longitudinal differential cross section ${d\sigma_{\textrm{\tiny L}}}/{dt}$ as a function of $Q^2$ (left panel) and $t$ (right panel).}
\end{figure}

Within the framework of the model used in this work, the longitudinal differential cross section ${d\sigma_{\textrm{\tiny L}}}/{dt}$ of the exclusive Sullivan process was calculated. The results are shown in figure \ref{fig_cross}. It can be seen that at low $t$ and $Q^2$, the obtained differential cross sections  ${d\sigma_{\textrm{\tiny L}}}/{dt}$ do not describe the experimental data well; however, with decreasing $t<-0.05~\textrm{GeV}^2$ and increasing $Q^2>1.0~ \textrm{GeV}^2$,  the experimental data are described better. The difficulties in describing the experimental data are supposedly related to the $\gamma\rho\pi$-vertex that was not taken into account in this investigation.

\section{Conclusions}
 In the present work, the physical properties of the pion on the on-shell and off-shell surfaces were investigated using the Bethe-Salpeter equation with the separable
 kernel of the quark-antiquark interaction. Two types of the vertex functions of the pion were compared: monopole and dipole. An optimal set of parameters ($m, \Lambda$) was selected for each function. Various pion constants were calculated ($f_{\pi} , r_{\pi\gamma}, <r^2_{\pi}>$, $\Gamma_{\pi^0 \rightarrow{} \gamma\gamma}$, $g(0,m_\pi^2)$), as well as the electromagnetic form factor $F_\pi(Q^2) $ on the on-shell surface and the transition form factor $F_{\pi\gamma}(Q^2)$ of the pion. The dipole vertex function turned out to be better than the monopole in comparison with experimental data. 

The half-off-shell electromagnetic form factors $F_1(Q^2,t)$ and $F_2(Q^2,t)$ were calculated. Additionally, the new form factor $g(Q^2,t)$  was obtained, which is nonzero on the mass shell and can be extracted from experimental data. This form factor was calculated using two methods: direct and using the Ward-Takahashi identity. The two calculations turned out to be the same, which implies the fulfillment of the Ward-Takahashi identity. Also, to fulfill the identity, it was necessary to take into account two contributions: the relativistic impulse approximation and the interaction current. 

It turned out that the off-shell factors $F_1(Q^2,t)$ and $g(Q^2,t)$ are insensitive to the choice of the vertex function at low $Q^2$ but differ at high $Q^2$. While the experimental data are not reproduced exactly, the calculated form factors exhibit the same behavior as the experimental data for any values of $Q^2$ and $t$. Also, the longitudinal differential
cross section $d\sigma_L/dt$ of the exclusive Sullivan process was calculated and was compared with experimental data. At low $t$, the calculated differential cross section differs from experimental data but at $t<-0.05~\textrm{GeV}^2$, calculations are consistent with experimental data.

It would be useful to consider the time-like part of the square of the transferred momentum ($q^2>0$) and calculate the contribution of the $\gamma\rho\pi$-vertex.

\begin{appendices}

\section{}\label{secA1}

Feynman parameterization is a method for evaluating loop integrals arising from Feynman diagrams with one or more loops.\\
The original formula:
\begin{equation}
    \frac{1}{a_1a_2...a_L}=(L-1)!\int_{0}^{1}\frac{\{dx\}_L}{(a_1x_1+...+a_Lx_L)^L},
\end{equation}
where $\{dx\}_L=dx_1...dx_L\delta(1-x_1-x_2-...-x_L)$.
The momentum quadratures used in the calculation have the form:
\begin{eqnarray}
&&F_L(D,l)=\frac{i}{\pi^2}\int\frac{dk}{[k^2-2lk+D]^L}=-\frac{(L-3)!}{(D-l^2+i\varepsilon)^{L-2}(L-1)!},\nonumber\\
&&F_L^{\mu}(D,l)=\frac{i}{\pi^2}\int\frac{k^{\mu}dk}{[k^2-2lk+D]^L}=l^{\mu}F_L(D,l),\nonumber\\
&&F_L^{\mu_1\mu_2}(D,l)=\frac{i}{\pi^2}\int\frac{k^{\mu_1}k^{\mu_2}dk}{[k^2-2lk+D]^L}=\{l^{\mu_1}l^{\mu_2}+\frac{(D-l^2)}{2(L-3)}g^{\mu_1\mu_2}\}F_L(D,l),\nonumber\\
&&F_L^{\mu_1\mu_2\mu_3}(D,l)=\frac{i}{\pi^2}\int\frac{k^{\mu_1}k^{\mu_2}k^{\mu_3}dk}{[k^2-2lk+D]^L}=\nonumber\\
&&=\{l^{\mu_1 }l^{\mu_2 }l^{\mu_3}+\frac{1}{2}\frac{(D-l^2)}{2 (L-3) }\sum_{i,j,k={1,3}}^{i\neq j\neq k}l^{\mu_{i} } {g}^{\mu_j \mu_k
   } \}F_L(D,l),\nonumber\\
   &&F_L^{\mu_1\mu_2\mu_3\mu_4}(D,l)=\frac{i}{\pi^2}\int\frac{k^{\mu_1}k^{\mu_2}k^{\mu_3}k^{\mu_4}dk}{[k^2-2lk+D]^L}=\nonumber\\
&&=\{l^{\mu_1}l^{\mu_2}l^{\mu_3}l^{\mu_4}+\frac{1}{4}\frac{(D-l^2)}{2(L-3)}\sum_{i,j,k,m={1,4}}^{i\neq j\neq k\neq m}g^{\mu_i\mu_j}l^{\mu_k}l^{\mu_m}+\nonumber\\
&&+\frac{1}{8}\frac{(D-l^2)^2}{4(L-4)(L-3)}\sum_{i,j,k,m={1,4}}^{i\neq j\neq k\neq m}g^{\mu_i\mu_j}g^{\mu_k\mu_m}
  \}F_L(D,l),\nonumber\\
 &&F_L^{{\mu_1\mu_2\mu_3\mu_4\mu_5}}(D,l)=\frac{i}{\pi^2}\int\frac{k^{\mu_1}k^{\mu_2}k^{\mu_3}k^{\mu_4}k^{\mu_5}dp}{[k^2-2lk+D]^L}=\nonumber\\
 &&=\{ l^{\mu_1}l^{\mu_2}l^{\mu_3}l^{\mu_4}l^{\mu_5} +\frac{1}{8}\frac{D-l^2}{2(L-3)}\sum_{i,j,k,m,n={1,5}}^{i\neq j\neq k\neq m\neq n}l^{\mu_i}l^{\mu_j}l^{\mu_k}g^{\mu_m\mu_n} +\nonumber\\
 &&+\frac{1}{12}\frac{(D-l^2)^2}{4(L-3)(L-4)}\sum_{i,j,k,m,n={1,5}}^{i\neq j\neq k\neq m\neq n}l^{\mu_i}g^{\mu_j\mu_k}g^{\mu_m\mu_n}\}F_L(D,l),\nonumber\\
 \label{feynformulas}
\end{eqnarray}
where the coefficients before the sum take into account the symmetry of the metric tensor $g^{\mu_1\mu_2}$.
\end{appendices}

\bibstyle{spphys}
\bibliography{references}

@article{Ito:1991sz,
    author = "Ito, Hiroshi and Buck, Warren and Gross, Franz",
    title = "{Current conservation and interaction currents with relativistic separable interactions}",
    reportNumber = "CEBAF-TH-90-11",
    doi = "10.1103/PhysRevC.43.2483",
    journal = "Phys. Rev. C",
    volume = "43",
    pages = "2483--2498",
    year = "1991"
}

@article{Ito:1991pv,
    author = "Ito, Hiroshi and Buck, W. W. and Gross, Franz",
    title = "{Electromagnetic properties of the pion as a composite Nambu-Goldstone boson}",
    reportNumber = "CEBAF-TH-91-12",
    doi = "10.1103/PhysRevC.45.1918",
    journal = "Phys. Rev. C",
    volume = "45",
    pages = "1918--1934",
    year = "1992"
}

@article{Anikin:1995cf,
    author = "Anikin, I. V. and Ivanov, Mikhail A. and Kulimanova, N. B. and Lyubovitskij, Valery E.",
    title = "{The Extended Nambu-Jona-Lasinio model with separable interaction: Low-energy pion physics and pion nucleon form-factor}",
    doi = "10.1007/BF01578675",
    journal = "Z. Phys. C",
    volume = "65",
    pages = "681--690",
    year = "1995"
}

@article{Nesterenko:1982gc,
    author = "Nesterenko, V. A. and Radyushkin, A. V.",
    title = "{Sum Rules and Pion Form-Factor in QCD}",
    reportNumber = "JINR-E2-82-204",
    doi = "10.1016/0370-2693(82)90528-7",
    journal = "Phys. Lett. B",
    volume = "115",
    pages = "410",
    year = "1982"
}

@article{Godfrey:1985xj,
    author = "Godfrey, S. and Isgur, Nathan",
    title = "{Mesons in a Relativized Quark Model with Chromodynamics}",
    doi = "10.1103/PhysRevD.32.189",
    journal = "Phys. Rev. D",
    volume = "32",
    pages = "189--231",
    year = "1985"
}

@inproceedings{Jacob:1988as,
    author = "Jacob, O. C. and Kisslinger, L. S.",
    title = "{Confining Bethe-Salpeter equation: A Light front formalism}",
    booktitle = "{3rd Lake Louise Winter Institute on QCD: Theory and Experiment}",
    year = "1988"
}

@article{Nambu:1961tp,
    author = "Nambu, Yoichiro and Jona-Lasinio, G.",
    editor = "Eguchi, T.",
    title = "{Dynamical Model of Elementary Particles Based on an Analogy with Superconductivity. 1.}",
    doi = "10.1103/PhysRev.122.345",
    journal = "Phys. Rev.",
    volume = "122",
    pages = "345--358",
    year = "1961"
}

@article{Maris:2000sk,
    author = "Maris, Pieter and Tandy, Peter C.",
    title = "{The pi, K+, and K0 electromagnetic form-factors}",
    eprint = "nucl-th/0005015",
    archivePrefix = "arXiv",
    reportNumber = "KSU-CNR-106-00",
    doi = "10.1103/PhysRevC.62.055204",
    journal = "Phys. Rev. C",
    volume = "62",
    pages = "055204",
    year = "2000"
}

@article{Bebek:1977pe,
    author = "Bebek, C. J. and others",
    title = "{Electroproduction of single pions at low epsilon and a measurement of the pion form-factor up to $q^2$ = 10-GeV$^2$}",
    reportNumber = "Print-77-0572 (HARVARD)",
    doi = "10.1103/PhysRevD.17.1693",
    journal = "Phys. Rev. D",
    volume = "17",
    pages = "1693",
    year = "1978"
}

@article{JeffersonLabFpi:2000nlc,
    author = "Volmer, J. and others",
    collaboration = "Jefferson Lab F(pi)",
    title = "{Measurement of the Charged Pion Electromagnetic Form-Factor}",
    eprint = "nucl-ex/0010009",
    archivePrefix = "arXiv",
    reportNumber = "JLAB-PHY-00-16",
    doi = "10.1103/PhysRevLett.86.1713",
    journal = "Phys. Rev. Lett.",
    volume = "86",
    pages = "1713--1716",
    year = "2001"
}

@article{JeffersonLabFpi-2:2006ysh,
    author = "Horn, T. and others",
    collaboration = "Jefferson Lab F(pi)-2",
    title = "{Determination of the Charged Pion Form Factor at Q**2 = 1.60 and 2.45-(GeV/c)**2}",
    eprint = "nucl-ex/0607005",
    archivePrefix = "arXiv",
    reportNumber = "JLAB-PHY-06-523",
    doi = "10.1103/PhysRevLett.97.192001",
    journal = "Phys. Rev. Lett.",
    volume = "97",
    pages = "192001",
    year = "2006"
}

@article{BaBar:2009rrj,
    author = "Aubert, Bernard and others",
    collaboration = "BaBar",
    title = "{Measurement of the gamma gamma* ---\ensuremath{>} pi0 transition form factor}",
    eprint = "0905.4778",
    archivePrefix = "arXiv",
    primaryClass = "hep-ex",
    reportNumber = "SLAC-PUB-13641, BABAR-PUB-09-006",
    doi = "10.1103/PhysRevD.80.052002",
    journal = "Phys. Rev. D",
    volume = "80",
    pages = "052002",
    year = "2009"
}

@article{Belle:2012wwz,
    author = "Uehara, S. and others",
    collaboration = "Belle",
    title = "{Measurement of $\gamma \gamma^* \to \pi^0$ transition form factor at Belle}",
    eprint = "1205.3249",
    archivePrefix = "arXiv",
    primaryClass = "hep-ex",
    reportNumber = "BELLE-PREPRINT-2012-16, KEK-PREPRINT-2012-8",
    doi = "10.1103/PhysRevD.86.092007",
    journal = "Phys. Rev. D",
    volume = "86",
    pages = "092007",
    year = "2012"
}

@article{CLEO:1997fho,
    author = "Gronberg, J. and others",
    collaboration = "CLEO",
    title = "{Measurements of the meson - photon transition form-factors of light pseudoscalar mesons at large momentum transfer}",
    eprint = "hep-ex/9707031",
    archivePrefix = "arXiv",
    reportNumber = "SLAC-PUB-9838, CLNS-97-1477, CLEO-97-7",
    doi = "10.1103/PhysRevD.57.33",
    journal = "Phys. Rev. D",
    volume = "57",
    pages = "33--54",
    year = "1998"
}

@article{CELLO:1990klc,
    author = "Behrend, H. J. and others",
    collaboration = "CELLO",
    title = "{A Measurement of the pi0, eta and eta-prime electromagnetic form-factors}",
    reportNumber = "DESY-90-110",
    doi = "10.1007/BF01549692",
    journal = "Z. Phys. C",
    volume = "49",
    pages = "401--410",
    year = "1991"
}

@article{Bernard:1986ti,
    author = "Bernard, V.",
    title = "{Remarks on Dynamical Breaking of Chiral Symmetry and Pion Properties in the Nambu and Jona-lasinio Model}",
    doi = "10.1103/PhysRevD.34.1601",
    journal = "Phys. Rev. D",
    volume = "34",
    pages = "1601--1605",
    year = "1986"
}

@article{Hatsuda:1985ey,
    author = "Hatsuda, T. and Kunihiro, T.",
    title = "{Critical Phenomena Associated with Chiral Symmetry Breaking and Restoration in QCD}",
    reportNumber = "KUNS-778",
    doi = "10.1143/PTP.74.765",
    journal = "Prog. Theor. Phys.",
    volume = "74",
    pages = "765",
    year = "1985"
}

@article{Gross:1991te,
    author = "Gross, Franz and Milana, Joseph",
    title = "{A Covariant, chirally symmetric, confining model of mesons}",
    reportNumber = "CEBAF-TH-90-09, WM-90-120",
    doi = "10.1103/PhysRevD.43.2401",
    journal = "Phys. Rev. D",
    volume = "43",
    pages = "2401--2417",
    year = "1991"
}

@article{Kekez:2020vfh,
    author = "Kekez, Dalibor and Klabu\v{c}ar, Dubravko",
    title = {{Pion observables calculated in Minkowski and Euclidean spaces with Ans\"atze for quark propagators}},
    eprint = "2006.02326",
    archivePrefix = "arXiv",
    primaryClass = "hep-ph",
    reportNumber = "ZTF-EP-20-03",
    doi = "10.1103/PhysRevD.107.094025",
    journal = "Phys. Rev. D",
    volume = "107",
    number = "9",
    pages = "094025",
    year = "2023"
}

@article{Hernandez-Pinto:2023yin,
    author = "Hern\'andez-Pinto, R. J. and Guti\'errez-Guerrero, L. X. and Bashir, A. and Bedolla, M. A. and Higuera-Angulo, I. M.",
    title = "{Electromagnetic form factors and charge radii of pseudoscalar and scalar mesons: A comprehensive contact interaction analysis}",
    eprint = "2301.11881",
    archivePrefix = "arXiv",
    primaryClass = "hep-ph",
    reportNumber = "JLAB-THY-23-3748",
    doi = "10.1103/PhysRevD.107.054002",
    journal = "Phys. Rev. D",
    volume = "107",
    number = "5",
    pages = "054002",
    year = "2023"
}

@article{Zhang:2024dhs,
    author = "Zhang, Jin-Li and Wu, Jun",
    title = "{Pion-photon and kaon-photon transition distribution amplitudes in the Nambu--Jona-Lasinio model}",
    eprint = "2402.12757",
    archivePrefix = "arXiv",
    primaryClass = "hep-ph",
    month = "2",
    year = "2024"
}

@article{ExtendedTwistedMass:2023hin,
    author = "Alexandrou, C. and others",
    collaboration = "Extended Twisted Mass",
    title = "{Pion transition form factor from twisted-mass lattice QCD and the hadronic light-by-light \ensuremath{\pi}0-pole contribution to the muon g-2}",
    eprint = "2308.12458",
    archivePrefix = "arXiv",
    primaryClass = "hep-lat",
    doi = "10.1103/PhysRevD.108.094514",
    journal = "Phys. Rev. D",
    volume = "108",
    number = "9",
    pages = "094514",
    year = "2023"
}

@article{Anikin:2000rq,
    author = "Anikin, I. V. and Dorokhov, A. E. and Tomio, L.",
    title = "{Pion structure in the instanton liquid model}",
    journal = "Phys. Part. Nucl.",
    volume = "31",
    pages = "509--537",
    year = "2000"
}

@article{Choi:2019nvk,
    author = "Choi, Ho-Meoyng and Frederico, T. and Ji, Chueng-Ryong and de Melo, J. P. B. C.",
    title = "{Pion off-shell electromagnetic form factors: data extraction and model analysis}",
    eprint = "1908.01185",
    archivePrefix = "arXiv",
    primaryClass = "hep-ph",
    reportNumber = "APCTP Pre2019-019, LFTC-19-9/47",
    doi = "10.1103/PhysRevD.100.116020",
    journal = "Phys. Rev. D",
    volume = "100",
    number = "11",
    pages = "116020",
    year = "2019"
}

@article{Leao:2024agy,
    author = "Le\~ao, Jurandi and de Melo, J. Pacheco B. C. and Frederico, T. and Choi, Ho-Meoyng and Ji, Chueng-Ryong",
    collaboration = "Jefferson Lab F\ensuremath{\pi}",
    title = "{Off-shell pion properties: Electromagnetic form factors and light-front wave functions}",
    eprint = "2406.07743",
    archivePrefix = "arXiv",
    primaryClass = "hep-ph",
    reportNumber = "LFTC - 24 -07/90",
    doi = "10.1103/PhysRevD.110.074035",
    journal = "Phys. Rev. D",
    volume = "110",
    number = "7",
    pages = "074035",
    year = "2024"
}

@article{Ward:1950xp,
    author = "Ward, John Clive",
    title = "{An Identity in Quantum Electrodynamics}",
    doi = "10.1103/PhysRev.78.182",
    journal = "Phys. Rev.",
    volume = "78",
    pages = "182",
    year = "1950"
}

@article{Takahashi:1957xn,
    author = "Takahashi, Y.",
    title = "{On the generalized Ward identity}",
    doi = "10.1007/BF02832514",
    journal = "Nuovo Cim.",
    volume = "6",
    pages = "371",
    year = "1957"
}

@article{Sullivan:1971kd,
    author = "Sullivan, J. D.",
    title = "{One pion exchange and deep inelastic electron - nucleon scattering}",
    doi = "10.1103/PhysRevD.5.1732",
    journal = "Phys. Rev. D",
    volume = "5",
    pages = "1732--1737",
    year = "1972"
}

@article{Rudy:1994qb,
    author = "Rudy, T. E. and Fearing, H. W. and Scherer, S.",
    title = "{The Off-shell electromagnetic form-factors of pions and kaons in chiral perturbation theory}",
    eprint = "hep-ph/9401302",
    archivePrefix = "arXiv",
    reportNumber = "TRI-PP-94-4",
    doi = "10.1103/PhysRevC.50.447",
    journal = "Phys. Rev. C",
    volume = "50",
    pages = "447--459",
    year = "1994"
}

@article{Gerardin:2019vio,
    author = "G\'erardin, Antoine and Meyer, Harvey B. and Nyffeler, Andreas",
    title = "{Lattice calculation of the pion transition form factor with $N_f=2+1$ Wilson quarks}",
    eprint = "1903.09471",
    archivePrefix = "arXiv",
    primaryClass = "hep-lat",
    reportNumber = "MITP-19-014",
    doi = "10.1103/PhysRevD.100.034520",
    journal = "Phys. Rev. D",
    volume = "100",
    number = "3",
    pages = "034520",
    year = "2019"
}

@article{Bondarenko:2025aep,
    author = "Bondarenko, S. G. and Slautin, M. K.",
    title = "{Pion in the Bethe-Salpeter Approach with Separable Kernel}",
    eprint = "2505.09421",
    archivePrefix = "arXiv",
    primaryClass = "hep-ph",
    doi = "10.1134/S1547477125700943",
    journal = "Phys. Part. Nucl. Lett.",
    volume = "22",
    number = "5",
    pages = "1000--1004",
    year = "2025"
}

@article{Bondarenko:2025qch,
    author = "Bondarenko, S. and Slautin, M.",
    title = "{Space-like pion off-shell form factors in the Bethe-Salpeter approach}",
    eprint = "2506.22153",
    archivePrefix = "arXiv",
    primaryClass = "hep-ph",
    month = "6",
    journal = "Memoirs of the Faculty of Physics Moscow Univ. /In Russian/",
    number = "6",
    pages = "2560102",
    year = "2025"
}

@book{Itzykson:1980rh,
    author = "Itzykson, C. and Zuber, J. B.",
    title = "{Quantum Field Theory}",
    isbn = "978-0-486-44568-7",
    publisher = "McGraw-Hill",
    address = "New York",
    series = "International Series In Pure and Applied Physics",
    year = "1980"
}

@article{Friesen:2025fhg,
    author = "Friesen, Alexandra and Kalinovsky, Yuriy and Khmelev, Alexander",
    title = "{Mesons in nonlocal model with four-dimensional separable kernel}",
    eprint = "2505.24542",
    archivePrefix = "arXiv",
    primaryClass = "hep-ph",
    doi = "10.1140/epja/s10050-025-01772-6",
    journal = "Eur. Phys. J. A",
    volume = "62",
    number = "1",
    pages = "6",
    year = "2026"
}

@article{PDG,
author = {Tanabashi, M and Grp, PD and Hagiwara, K and Hikasa, K and Nakamura, K and Sumino, Yukinari and Takahashi, F and Tanaka, Junta and Agashe, K and Aielli, Giulio and Amsler, Claude and Antonelli, Mario and Asner, DM and Baer, Howard and Banerjee, S and Barnett, RM and Basaglia, T and Bauer, Christian and Schaffner, P},
year = {2018},
month = {08},
pages = {},
title = {Review of Particle Physics: Particle Data Group},
volume = {98},
journal = {Physical Review D}
}

@article{JeffersonLab:2008jve,
    author = "Huber, G. M. and others",
    collaboration = "Jefferson Lab",
    title = "{Charged pion form-factor between Q**2 = 0.60-GeV**2 and 2.45-GeV**2. II. Determination of, and results for, the pion form-factor}",
    eprint = "0809.3052",
    archivePrefix = "arXiv",
    primaryClass = "nucl-ex",
    reportNumber = "JLAB-PHY-08-864",
    doi = "10.1103/PhysRevC.78.045203",
    journal = "Phys. Rev. C",
    volume = "78",
    pages = "045203",
    year = "2008"
}

@article{JeffersonLab:2008gyl,
    author = "Blok, H. P. and others",
    collaboration = "Jefferson Lab",
    title = "{Charged pion form factor between $Q^2$=0.60 and 2.45 GeV$^2$. I. Measurements of the cross section for the ${^1}$H($e,e'\pi^+$)$n$ reaction}",
    eprint = "0809.3161",
    archivePrefix = "arXiv",
    primaryClass = "nucl-ex",
    reportNumber = "JLAB-PHY-08-873",
    doi = "10.1103/PhysRevC.78.045202",
    journal = "Phys. Rev. C",
    volume = "78",
    pages = "045202",
    year = "2008"
}

@article{JeffersonLabFpi:2007vir,
    author = "Tadevosyan, V. and others",
    collaboration = "Jefferson Lab F(pi)",
    title = "{Determination of the pion charge form-factor for Q**2 = 0.60-GeV**2 - 1.60-GeV**2}",
    eprint = "nucl-ex/0607007",
    archivePrefix = "arXiv",
    reportNumber = "JLAB-PHY-06-522",
    doi = "10.1103/PhysRevC.75.055205",
    journal = "Phys. Rev. C",
    volume = "75",
    pages = "055205",
    year = "2007"
}

@article{Zhang:2025iiw,
    author = "Zhang, Jin-Li",
    title = "{The Kaon Off-Shell Generalized Parton Distributions and Transverse Momentum Dependent Parton Distributions}",
    eprint = "2509.01854",
    archivePrefix = "arXiv",
    primaryClass = "hep-ph",
    reportNumber = "Particles 2025, 8(4), 85",
    doi = "10.3390/particles8040085",
    journal = "Particles",
    volume = "8",
    number = "4",
    pages = "85",
    year = "2025"
}

\end{document}